\documentstyle[twoside]{article}
\catcode`\@=11
\long\def\@makefntext#1{
\protect\noindent \hbox to 3.2pt {\hskip-.9pt  
$^{{\eightrm\@thefnmark}}$\hfil}#1\hfill}		

\def\thefootnote{\fnsymbol{footnote}}
\def\@makefnmark{\hbox to 0pt{$^{\@thefnmark}$\hss}}	
	
\def\ps@myheadings{\let\@mkboth\@gobbletwo
\def\@oddhead{\hbox{}
\rightmark\hfil\eightrm\thepage}   
\def\@oddfoot{}\def\@evenhead{\eightrm\thepage\hfil
\leftmark\hbox{}}\def\@evenfoot{}
\def\sectionmark##1{}\def\subsectionmark##1{}}

\oddsidemargin=\evensidemargin
\renewcommand{\thefootnote}{\fnsymbol{footnote}}

\newcounter{sectionc}\newcounter{subsectionc}\newcounter{subsubsectionc}
\renewcommand{\section}[1] {\vspace{12pt}
\setcounter{subsectionc}{0}\setcounter{subsubsectionc}{0}\noindent 
	{\tenbf\thesectionc. #1}\par\vspace{5pt}}
\renewcommand{\subsection}[1] {\vspace{12pt}\addtocounter{subsectionc}{1} 
	\setcounter{subsubsectionc}{0}\noindent 
	{\bf\thesectionc.\thesubsectionc. {\kern1pt \bfit #1}}\par\vspace{5pt}}
\renewcommand{\subsubsection}[1] {\vspace{12pt}\addtocounter{subsubsectionc}{1}
	\noindent{\tenrm\thesectionc.\thesubsectionc.\thesubsubsectionc.
	{\kern1pt \tenit #1}}\par\vspace{5pt}}
\newcommand{\nonumsection}[1] {\vspace{12pt}\noindent{\tenbf #1}
	\par\vspace{5pt}}

\newcounter{appendixc}
\newcounter{subappendixc}[appendixc]
\newcounter{subsubappendixc}[subappendixc]
\renewcommand{\thesubappendixc}{\Alph{appendixc}.\arabic{subappendixc}}
\renewcommand{\thesubsubappendixc}
	{\Alph{appendixc}.\arabic{subappendixc}.\arabic{subsubappendixc}}

\renewcommand{\appendix}[1] {\vspace{12pt}
        \refstepcounter{appendixc}
        \setcounter{figure}{0}
        \setcounter{table}{0}
        \setcounter{lemma}{0}
        \setcounter{theorem}{0}
        \setcounter{corollary}{0}
        \setcounter{definition}{0}
        \setcounter{equation}{0}
        \renewcommand{\thefigure}{\Alph{appendixc}.\arabic{figure}}
        \renewcommand{\thetable}{\Alph{appendixc}.\arabic{table}}
        \renewcommand{\theappendixc}{\Alph{appendixc}}
        \renewcommand{\thelemma}{\Alph{appendixc}.\arabic{lemma}}
        \renewcommand{\thetheorem}{\Alph{appendixc}.\arabic{theorem}}
        \renewcommand{\thedefinition}{\Alph{appendixc}.\arabic{definition}}
        \renewcommand{\thecorollary}{\Alph{appendixc}.\arabic{corollary}}
        \renewcommand{\theequation}{\Alph{appendixc}.\arabic{equation}}
        \noindent{\tenbf Appendix \theappendixc #1}\par\vspace{5pt}}
\newcommand{\subappendix}[1] {\vspace{12pt}
        \refstepcounter{subappendixc}
        \noindent{\bf Appendix \thesubappendixc. {\kern1pt \bfit #1}}
	\par\vspace{5pt}}
\newcommand{\subsubappendix}[1] {\vspace{12pt}
        \refstepcounter{subsubappendixc}
        \noindent{\rm Appendix \thesubsubappendixc. {\kern1pt \tenit #1}}
	\par\vspace{5pt}}

\newcommand{\textlineskip}{\baselineskip=13pt}
\newcommand{\smalllineskip}{\baselineskip=10pt}

\def\eightcirc{
\begin{picture}(0,0)
\put(4.4,1.8){\circle{6.5}}
\end{picture}}
\def\eightcopyright{\eightcirc\kern2.7pt\hbox{\eightrm c}}

\def\abstracts#1#2#3{{
	\centering{\begin{minipage}{4.5in}\baselineskip=10pt\footnotesize
	\parindent=0pt #1\par 
	\parindent=15pt #2\par
	\parindent=15pt #3
	\end{minipage}}\par}} 

\renewenvironment{thebibliography}[1]
	{\frenchspacing
	 \ninerm\baselineskip=11pt
	 \begin{list}{\arabic{enumi}.}
	{\usecounter{enumi}\setlength{\parsep}{0pt}
	 \setlength{\leftmargin 12.7pt}{\rightmargin 0pt} 
	 \setlength{\itemsep}{0pt} \settowidth
	{\labelwidth}{#1.}\sloppy}}{\end{list}}

\newcounter{itemlistc}
\newcounter{romanlistc}
\newcounter{alphlistc}
\newcounter{arabiclistc}

\newcommand{\fcaption}[1]{
        \refstepcounter{figure}
        \setbox\@tempboxa = \hbox{\footnotesize Fig.~\thefigure. #1}
        \ifdim \wd\@tempboxa > 5in
           {\begin{center}
        \parbox{5in}{\footnotesize\smalllineskip Fig.~\thefigure. #1}
            \end{center}}
        \else
             {\begin{center}
             {\footnotesize Fig.~\thefigure. #1}
              \end{center}}
        \fi}

\newcommand{\tcaption}[1]{
        \refstepcounter{table}
        \setbox\@tempboxa = \hbox{\footnotesize Table~\thetable. #1}
        \ifdim \wd\@tempboxa > 5in
           {\begin{center}
        \parbox{5in}{\footnotesize\smalllineskip Table~\thetable. #1}
            \end{center}}
        \else
             {\begin{center}
             {\footnotesize Table~\thetable. #1}
              \end{center}}
        \fi}

\def\@citex[#1]#2{\if@filesw\immediate\write\@auxout
	{\string\citation{#2}}\fi
\def\@citea{}\@cite{\@for\@citeb:=#2\do
	{\@citea\def\@citea{,}\@ifundefined
	{b@\@citeb}{{\bf ?}\@warning
	{Citation `\@citeb' on page \thepage \space undefined}}
	{\csname b@\@citeb\endcsname}}}{#1}}

\newif\if@cghi
\def\cite{\@cghitrue\@ifnextchar [{\@tempswatrue
	\@citex}{\@tempswafalse\@citex[]}}
\def\citelow{\@cghifalse\@ifnextchar [{\@tempswatrue
	\@citex}{\@tempswafalse\@citex[]}}
\def\@cite#1#2{{$\null^{#1}$\if@tempswa\typeout
	{IJCGA warning: optional citation argument 
	ignored: `#2'} \fi}}

\def\pmb#1{\setbox0=\hbox{#1}
	\kern-.025em\copy0\kern-\wd0
	\kern.05em\copy0\kern-\wd0
	\kern-.025em\raise.0433em\box0}

\def\fnt#1#2{\footnotetext{\kern-.3em
	{$^{\mbox{\scriptsize #1}}$}{#2}}}

\def\fpage#1{\begingroup
\voffset=.3in
\thispagestyle{empty}\begin{table}[b]\centerline{\footnotesize #1}
	\end{table}\endgroup}

\def\runninghead#1#2{\pagestyle{myheadings}
\markboth{{\protect\footnotesize\it{\quad #1}}\hfill}
{\hfill{\protect\footnotesize\it{#2\quad}}}}
\font\tenrm=cmr10
\font\tenit=cmti10 
\font\tenbf=cmbx10
\font\bfit=cmbxti10 at 10pt
\font\ninerm=cmr9

\font\eightrm=cmr8

\def\qed{\hbox{${\vcenter{\vbox{			
   \hrule height 0.4pt\hbox{\vrule width 0.4pt height 6pt
   \kern5pt\vrule width 0.4pt}\hrule height 0.4pt}}}$}}

\renewcommand{\thefootnote}{\fnsymbol{footnote}}	

\newcommand{\ccopyrightheading}[1]
	{\vspace*{-2.5cm}\smalllineskip{\flushleft
	{\footnotesize A review to appear in the International Journal of Modern Physics
A #1}\\
	{\footnotesize \ }\\
	 }}
\begin{document}

\runninghead{} {}
\thispagestyle{empty}

\normalsize\textlineskip
\thispagestyle{empty}
\setcounter{page}{1}

\ccopyrightheading{}			

\vspace*{0.88truein}

\fpage{1}
\centerline{\bf BPS MONOPOLES}
\vspace*{0.37truein}
\centerline{\footnotesize PAUL M. SUTCLIFFE\footnote{Email\  P.M.Sutcliffe@ukc.ac.uk}}
\vspace*{0.015truein}
\centerline{\footnotesize\it Institute of Mathematics, University of Kent at Canterbury, }
\baselineskip=10pt
\centerline{\footnotesize\it Canterbury CT2 7NZ, England}
\vspace*{0.225truein}
\ \\ \ \\ \ \\
\vspace*{0.21truein}
\abstracts{We review classical BPS monopoles, their moduli spaces, twistor descriptions
and dynamics.
Particular emphasis is placed upon symmetric monopoles, where recent progress
 has been made. Some remarks on the role of monopoles
in S-duality and Seiberg-Witten theory are also made.
}{}{}



\textheight=7.8truein
\setcounter{footnote}{0}
\renewcommand{\thefootnote}{\alph{footnote}}


\input epsf
\renewcommand{\theequation}{\arabic{section}.\arabic{equation}}
\newcommand{\news}{\setcounter{equation}{0}\addtocounter{section}{1}\addtocounter{sectionc}{1}}
\newcommand{\be}{\begin{equation}}
\newcommand{\ee}{\end{equation}}
\newcommand{\bea}{\begin{eqnarray}}
\newcommand{\eea}{\end{eqnarray}}
\newcommand{\bean}{\begin{eqnarray*}}
\newcommand{\eean}{\end{eqnarray*}}
\font\upright=cmu10 
\font\sans=cmss10
\newcommand{\ssf}{\sans}
\newcommand{\stroke}{\vrule height6.5pt width0.4pt depth-0.1pt}
\newcommand{\Z}{\hbox{\upright\rlap{\ssf Z}\kern 2.7pt {\ssf Z}}}
\newcommand{\ZZ}{\Z\hskip -10pt \Z_2}
\newcommand{\C}{{\rlap{\rlap{C}\kern 3.0pt\stroke}\phantom{C}}}
\newcommand{\R}{\hbox{\upright\rlap{I}\kern 1.7pt R}}
\newcommand{\CP}{\C{\upright\rlap{I}\kern 1.5pt P}}
\newcommand{\half}{\frac{1}{2}}
\newcommand{\mt}{\rlap{\ssf T}\kern 3.0pt{\ssf T}}
\newcommand{\spc}{spectral curve }
\newcommand{\ode}{{\scriptsize ODE}}
\newcommand{\ivp}{{\scriptsize IVP}}
\newcommand{\identity}{{\upright\rlap{1}\kern 2.0pt 1}}
\newcommand{\bm}{\boldmath}
\newcommand{\alim}{3^{-5/4}\sqrt{2}}
\newcommand{\tr}{\vert\vert\Phi\vert\vert}
\newcommand{\wpp}{{\wp^\prime}}
\newcommand{\rhobf}{\mbox{\boldmath $\rho$}}

\news
\vspace*{1pt}\textlineskip	
\section{Introduction}	
\vspace*{-0.5pt}
\noindent
Monopoles are topological soliton solutions in three space dimensions,
 which arise in Yang-Mills-Higgs
gauge theories where the non-abelian gauge group $G$ is spontaneously broken
by the Higgs field to a residual symmetry group $H$. The Higgs field at infinity
defines a map from $S^2$ to the coset space of vacua $G/H$, so if 
$\pi_2(G/H)$ is non-trivial then all solutions have a topological characterization.
The simplest case is to take $G=SU(2)$ broken to
$H=U(1)$ by a Higgs field in the adjoint representation. We shall concentrate
upon this case and indicate how the methods and results extend to higher rank
gauge groups in Section 5.

The Lagrangian density, with the usual symmetry breaking potential is
\be
{\cal L}=\frac{1}{8}\mbox{tr}(F_{\mu\nu}F^{\mu\nu})-\frac{1}{4}\mbox{tr}
(D_\mu\Phi D^\mu\Phi)
-\frac{1}{8}\lambda(\tr^2-1)^2
\label{lag}
\ee
where $\tr^2=-\frac{1}{2}\mbox{tr}\Phi^2$ is the square of the length of the Higgs field.
In the $SU(2)$ case we have that $\pi_2(SU(2)/U(1))=\Z$, so that each configuration
has an associated topological integer, or winding number, $k$, which may be
expressed as
\be
k=\frac{1}{8\pi}\int \mbox{tr}(B_iD_i\Phi)\ d^3{\bf x}.
\ee
Here $B_i$ denotes the magnetic part of the gauge field, $B_i=\half\epsilon_{ijk}F_{jk}$,
and indices run over the spatial values $1,2,3.$
The integer $k$ is the degree of the map given by the Higgs field at infinity
\be
\widehat \Phi: S^2\mapsto S^2
\ee
and is known as the monopole number, or charge,
 since it determines the magnetic charge of the
solution as follows. The $U(1)$ residual symmetry group is identified as the
group of electromagnetism, and the only component of the non-abelian gauge
field which survives at infinity is the one in the direction of $\widehat \Phi$, 
 the asymptotic Higgs field.
This allows us to define abelian magnetic and electric fields
\be
b_i=\half\mbox{tr}(B_i\widehat \Phi), \hskip 1cm 
e_i=\half\mbox{tr}(E_i\widehat \Phi)
\ee
and it can be shown, by using Stokes' theorem, that the
associated magnetic charge is exactly $4\pi k.$ 

In order to make progress with the classical solutions of the second order field
equations which follow from (\ref{lag}) it is helpful to consider the BPS limit\cite{Bo,PS}, 
in which the Higgs potential is removed by setting $\lambda=0.$
We shall consider only this case from now on. Then, as Bogomolny pointed out\cite{Bo},
by completing a square in the energy integral 
\be
E=\frac{1}{4}\int
-\mbox{tr}(D_i\Phi D_i\Phi+ B_iB_i+E_iE_i+D_t\Phi D_t\Phi)
\ d^3{\bf x}
\ee
a bound on $E$, in terms of the monopole number  $k$, can be obtained 
\be
E\ge 4\pi\vert k\vert.
\label{bound}
\ee
Moreover, all solutions of the second order equations which attain this bound
 are static solutions that solve the first order Bogomolny equation
\be
D_{i}\Phi=-B_i
\label{bog}
\ee
(or the one obtained from above by a change of sign for $k<0$).

For solutions of the Bogomolny equation (\ref{bog}) the energy may be expressed
in the convenient form\cite{Wa1}
\be
 E=\half\int \partial_i\partial_i \|\Phi\|^2 \ d^3{\bf x}
\label{lap}
\ee
allowing the energy density to be computed from knowledge of the Higgs field alone.

The charge one solution of equation (\ref{bog}) has a spherically symmetric form
and was first written down by Prasad \& Sommerfield\cite{PS}. It is given by
\be
\Phi=ix_j\sigma_j\frac{r\mbox{cosh} r-\mbox{sinh} r }{r^2 \mbox{sinh} r },
 \hskip 1cm
A_i=-i\epsilon_{ijk}\sigma_jx_k\frac{\mbox{sinh} r-r}{r^2 \mbox{sinh} r}.
\label{k1}
\ee
This monopole is positioned at the origin, but clearly three parameters can be
introduced into the solution by a translation which places the monopole at
an arbitrary location in $\R^3.$ A further parameter can also be introduced
into this solution, but its appearance is more subtle. Consider  gauge
transformations of the form $g=\mbox{exp}(\chi \Phi)$, which is a zero mode
since the potential energy is independent of the constant $\chi.$ 
However, if one returns
to the second order field equations then it turns out\cite{JZ} that a 
 transformation of this form, but where $\chi$ is now linearly time dependent,
 also gives a solution,
and now the monopole becomes a dyon, with  electric charge proportional
to the rate of change of $\chi.$ Thus it is useful to include the phase
$\chi$ as one of the moduli in the solutions of the Bogomolny equation (\ref{bog}).
Hence the moduli space, ${\cal M}_1$, of charge one solutions is 4-dimensional.
Mathematically, to correctly define this 4-dimensional moduli space of gauge
inequivalent solutions one must carefully state the type of gauge transformations
which are allowed, and consider framed monopoles and transformations\cite{AH}.

In the BPS limit of a massless Higgs there is now both a long-range magnetic repulsion
and a long-range scalar attraction between the fields of two well-separated and equally
charged monopoles.
By studying the field equations Manton\cite{Ma4} was able to show that these long-range forces
exactly cancel, thus providing for the possibility of static multi-monopoles ie. solutions
of (\ref{bog}) with $k>1.$ 

Multi-monopole solutions indeed exist and are the topic of discussion in Section 2.
Briefly, it is now known that the moduli space of charge $k$ solutions, ${\cal M}_k$, is
a $4k$-dimensional manifold\cite{We1,CG,Do,JT}. Roughly speaking, if the monopoles are all
well-separated then the $4k$ parameters represent three position coordinates and a phase for
each of the $k$ monopoles. However, for points in ${\cal M}_k$ which describe monopoles that
are close together, we shall see that the situation is more interesting. The explicit construction
of any multi-monopole solution turned out to be a difficult task, which required the
application of powerful twistor methods, as we now review.

\news
\vspace*{1pt}\textlineskip	
\section{Twistor Methods}
\vspace*{-0.5pt}
\noindent
As a first attempt to construct multi-monopole solutions one might try to generalize
the spherically symmetric $k=1$ solution (\ref{k1}). However, Bogomolny showed\cite{Bo}
that this is in fact the unique spherically symmetric solution, so there are no
spherically symmetric monopoles for $k>1.$ This makes the task of explicitly constructing any 
multi-monopole solution more difficult,
and indeed it would probably be impossible if it were not for the fact that the Bogomolny
equation (\ref{bog}) has the special property of being integrable, as we explain in the following.

Ward's twistor transform for self-dual Yang-Mills gauge fields\cite{Wa4,WaW} relates
instanton solutions in $\R^4$ to certain holomorphic vector bundles over the standard 
complex 3-dimensional twistor space \CP$^3.$ By constructing such bundles the self-dual
gauge fields can be extracted and this can be done explicitly in a number of cases\cite{AW}.
Now to be able to apply this technique to monopoles we need to find the connection between
self-dual gauge fields in $\R^4$ and monopoles, which are objects in $\R^3.$

The observation of Manton\cite{Ma5} is that if one considers the self-dual equation
\be
F_{\mu\nu}=\half\epsilon_{\mu\nu\alpha\beta}F_{\alpha\beta}
\label{sdym}
\ee
for the gauge field in four dimensions (so greek indices take the values $1,2,3,4$)
 and dimensionally reduces, by
setting all functions independent of the $x_4$ coordinate, then (\ref{sdym}) reduces
to the Bogomolny equation (\ref{bog}) after the identification $A_4=\Phi.$
Thus monopoles may be thought of as some particular kinds of self-dual gauge fields, though
they are not instantons, since they are required to have infinite action in order to have the
required $x_4$ dependence. For the particular case of the charge one monopole (\ref{k1})
the solution can be obtained using the form of the 
Corrigan-Fairlie-t'Hooft-Wilczek ansatz\cite{CF,tH,Wil} which gives a subset of self-dual gauge
fields, though no multi-monopoles can be obtained in this way\cite{Ma5}.

Knowing the form of the $k=1$ monopole solution, one can construct the associated vector
bundle to which it corresponds and then hope to generalize this to obtain the vector
bundle of a $k=2$ solution and hence a charge two monopole. Ward's original description
of this procedure\cite{Wa1} was in terms of bundles over \CP$^3$, which have
a particular special form to obtain the required $x_4$ independence of the gauge fields.
However, as later described by Hitchin\cite{Hi1}, the dimensional reduction can be made 
at the twistor level too, to obtain a direct correspondence between monopoles and
bundles over the mini-twistor space \mt, which is a 2-dimensional
complex manifold isomorphic to the holomorphic tangent bundle to the
Riemann sphere T\CP$^1$. It is helpful in connecting with other approaches if we adopt
this reduced description.

To put coordinates on \mt\ let $\zeta$ be the standard inhomogeneous coordinate on the
base space and $\eta$ the complex fibre coordinate. For the twistor transform these
twistor coordinates are related to the space coordinates $x_1,x_2,x_3$ via the relation
\be
\eta=\frac{(x_1+ix_2)}{2}-x_3\zeta-\frac{(x_1-ix_2)}{2}\zeta^2.
\label{fibre}
\ee
Monopoles correspond to certain rank two vector bundles over \mt, which may be characterized
by a  $2\times 2$ patching matrix which relates the local trivializations over the two patches 
$U_1=\{\zeta:\vert \zeta\vert \le 1\}$
and $U_2=\{\zeta:\vert \zeta\vert \ge 1\}.$ For charge $k$ monopoles the patching matrix, $F$, 
may be taken to have the Atiyah-Ward form\cite{AW}
\be
F=\left(\begin{array}{cc}
\zeta^k & \Gamma \\
0 & \zeta^{-k} \\
\end{array}\right).
\label{AW}
\ee
To extract the gauge fields from the bundle over twistor space requires the patching
matrix to be \lq split\rq\ as $F=H_2 H_1^{-1}$ on the overlap $U_1\cap U_2$,
 where $H_1$ and $H_2$ are regular and holomorphic 
in the patches $U_1$ and $U_2$ respectively.

For a patching matrix of the Atiyah-Ward form this \lq splitting\rq\ can be done by a contour
integral. From the Taylor-Laurent coefficients
\be
\Delta_p=\frac{1}{2\pi i}\oint_{\vert\zeta\vert =1}\Gamma \zeta^{p-1} \ d\zeta
\label{conint}
\ee
the gauge fields can be computed. For example there is the elegant formula\cite{Pr2}
\be
\|\Phi\|^2=1-\partial_i\partial_i\log D
\label{elegant}
\ee
where $D$ is the determinant of the $k\times k$ banded matrix with entries
\be
D_{pq}=\Delta_{p+q-k-1}, \hskip 1cm 1\le p,q\le k.
\ee

For charge $k$ monopoles the function $\Gamma$ in the Atiyah-Ward ansatz (\ref{AW})
has the form\cite{WaW,CG}
\be
\Gamma=\frac{\zeta^k}{S}
(e^{(-x_1+ix_2)\zeta-x_3}+(-1)^ke^{(-x_1-ix_2)\zeta^{-1}+x_3})
\ee
where $S$ is a polynomial in $\eta$ of degree $k$, with coefficients which are
polynomials in $\zeta.$ 

The $k=1$ solution (\ref{k1}) is obtain by taking $S=\eta.$ In this case the
contour integral (\ref{conint}) gives
\be
\Delta_0=\frac{2\mbox{sinh} r}{r}
\ee
and since $k=1$ then $D=\Delta_0$ and (\ref{elegant}) gives
\be
\|\Phi\|^2=1-\partial_i\partial_i\log \frac{2\mbox{sinh} r}{r} =
\frac{(r\mbox{cosh} r-\mbox{sinh} r)^2}{(r\mbox{sinh} r)^2} 
\ee
which clearly agrees with (\ref{k1}).
Similarly, by taking 
\be
S=\eta-\frac{(a_1+ia_2)}{2}+a_3\zeta+\frac{(a_1-ia_2)}{2}\zeta^2
\label{star}
\ee
a monopole with position ${\bf x}=(a_1,a_2,a_3)$ is obtained.

Ward\cite{Wa1} was able
to generalize the $k=1$ solution to present a 2-monopole solution which
 is given by taking
\be
S=\eta^2+\frac{\pi^2}{4}\zeta^2.
\label{sc2}
\ee
As stated earlier, it was known that such a 
solution could not be spherically symmetric and in fact Ward's 2-monopole solution has
an axial symmetry, so that a surface of constant energy density is a torus. The traditional
definition of the position of a monopole is taken to be where the Higgs field is zero. This
toroidal monopole has a double zero at the origin and no others, so this configuration
may be thought of as two monopoles both of which are located at the origin.

For the general 2-monopole solution there is one important parameter, related to the
separation of the monopoles, with the other seven parameters being accounted for by
the position of the centre of mass, an overall phase and spatial $SO(3)$ rotations of
the whole configuration. This one-parameter family of solutions was also
constructed by Ward\cite{Wa2} and corresponds to the function
\be
S=\eta^2-\frac{K^2}{4}(m+2(m-2)\zeta^2+m\zeta^4)
\label{sc2b}
\ee
where $m\in[0,1)$ and $K$ is the complete elliptic integral of the first
kind with parameter $m$. If $m=0$ then $K=\pi/2$ and (\ref{sc2b}) reduces to
(\ref{sc2}), representing coincident monopoles. In the limit as $m\rightarrow 1$
then $K\rightarrow\infty$ and (\ref{sc2b}) becomes asymptotic to the product
\be
S=(\eta+\frac{K}{2}(1-\zeta^2))(\eta-\frac{K}{2}(1-\zeta^2))
\ee
which, by comparison with (\ref{star}), can be seen to describe two well-separated
monopoles with positions $(\pm K,0,0).$

It should be noted that at around the same time that Ward produced his
2-monopole solutions using twistor methods, a more traditional integrable
systems approach was taken by Forg\'acs, Horv\'ath \& Palla\cite{FHP} and the same
results obtained. This method makes use of the fact that the Bogomolny
equation (\ref{bog}) (in a suitable formulation) can be written as the
compatibility condition of an overdetermined linear system. The linear system
can be solved in terms of projectors and the corresponding gauge fields
extracted.

An axially symmetric monopole exists for all $k>1$, with the functions
$S$ that generalize the $k=2$ example (\ref{sc2}) being found by Prasad \& Rossi\cite{PR,Pr1} 
to be
\be
S=\prod_{l=0}^g\{\eta^2+(l+\half)^2\pi^2\zeta^2\}
 \hskip 0.5cm  \mbox{for} \hskip 0.5cm k=2g+2 
\label{scaxeven}
\ee
\be
S=\eta\prod_{l=1}^g\{\eta^2+l^2\pi^2\zeta^2\}
 \hskip 1.5cm  \mbox{for} \hskip 0.5cm k=2g+1
\label{scaxodd}
\ee

A main difficulty with this direct twistor approach is finding the polynomials
 $S$ so that a non-singular solution of the Bogomolny equation (\ref{bog}) is
obtained. By explicit computation Ward was able to show that the 
axially symmetric 2-monopole solution corresponding to (\ref{sc2}) is smooth,
and this can then be used to show that, at least for solutions corresponding to
 (\ref{sc2b}) which are close to this one, then they too are smooth.
A general discussion of non-singularity was given by Hitchin\cite{Hi1}
and we shall return to this shortly.

An analysis of the degrees of freedom in the function $S$ together
with the constraints of reality and non-singularity, allowed Corrigan \& Goddard\cite{CG}
 to deduce that a charge $k$ monopole solution has $4k$ degrees of freedom,
as we have mentioned earlier. However, for $k>2$ very few explicit examples of functions
$S$ are known, although some have recently been obtained by considering particularly
symmetric cases (see later). Note that even if the function $S$ is known it is
still a difficult task to perform the contour integrals (\ref{conint}) and
extract the gauge fields explicitly, with the tractable cases only likely to be
those in which $S$ factors into a product containing no greater than quadratic
polynomials in $\eta.$

We have seen that a monopole is determined by a vector bundle over \mt\ and
that this in turn is determined by a polynomial $S$. In fact it turns out
that this polynomial occurs in many of the different twistor descriptions
of monopoles and is a central object.
A more direct relation between this polynomial and the monopole was introduced by
Hitchin\cite{Hi1} and we shall now review this topic of spectral curves.

In Hitchin's approach \mt\ is identified with the space of directed lines
in $\R^3.$ The base space coordinate $\zeta$ defines a direction in $\R^3$ (via
the usual Riemann sphere description of $S^2$) and the fibre $\eta$ is a complex
coordinate in a plane orthogonal to this line.
Given a line in $\R^3$, determined by a  point in \mt,
 one then considers the linear
differential equation
\be
(D_u-i\Phi)v=0
\label{hitchins}
\ee
for the complex doublet $v$, where $u$ is the coordinate along the line and
$D_u$ denotes the covariant derivative in the direction of the line.
 This equation has two 
independent solutions and a basis $(v_0,v_1)$  can be
chosen such that 
\begin{eqnarray} \lim_{u\rightarrow
    \infty}v_0(u)u^{\;-k/2}e^{u}&=&e_0,\\
   \lim_{u\rightarrow
    \infty}v_1(u)u^{\;k/2}e^{-u}&=&e_1\nonumber\end{eqnarray}
where $e_0$, $e_1$ are constant in some asymptotically flat
gauge. Thus $v_0$ is bounded and $v_1$ is unbounded as
$u\rightarrow\infty$. Clearly a similar description exists in terms of
a basis $(v_0',v_1')$ of bounded and unbounded solutions in the opposite direction as
$u\rightarrow -\infty.$ Now the line along which we consider Hitchin's equation
(\ref{hitchins}) is called a spectral line if the solution is decaying
in both directions $u\rightarrow\pm\infty$ ie. the unbounded component
of the solution is absent as we move off to infinity in either direction
along the line. The set of all spectral lines defines a curve of genus 
$(k-1)^2$ in \mt\
called the spectral curve, which for a charge $k$ monopole has the form\cite{Hi1}
\be
S=\eta^k+\eta^{k-1} a_1(\zeta)+\ldots+\eta^r a_{k-r}(\zeta)
+\ldots+\eta a_{k-1}(\zeta)+a_k(\zeta)=0
\label{gencurve}
\ee
where, for $1\leq r\leq k$, $a_r(\zeta)$ is a polynomial in $\zeta$ of
maximum degree $2r$.
However, general curves of this form will only correspond 
to $k$-monopoles if they
satisfy the reality condition
\be 
a_r(\zeta)=
(-1)^r\zeta^{2r}\overline{a_r(-1/\overline{\zeta})}
\label{reality}
\ee
and some difficult non-singularity conditions\cite{Hi1}.

This function $S$, whose zero set gives the spectral curve, is precisely
the function which occurred earlier in the determination of the Atiyah-Ward patching
matrix. Moreover, Hitchin\cite{Hi1} was able to prove  that all monopole solutions
can be constructed from the Atiyah-Ward class of bundles using Ward's method.
By an analysis of the spectral curve singularity constraints Hurtubise\cite{Hu1}
was able to derive the spectral curve (\ref{sc2b}) of the general 2-monopole
solution.

Although the spectral curve has now appeared in two approaches the main problem
still lies in satisfying the difficult non-singularity conditions which
a general curve must satisfy in order to be a spectral curve and thus correspond
to a monopole. A third approach was introduced by Nahm\cite{Na} and this has
the great advantage that non-singularity is manifest, although there is a price to
be paid for this, which is that the transform requires the solution of a matrix
nonlinear ordinary differential equation.

The Atiyah-Drinfeld-Hitchin-Manin (ADHM) construction\cite{ADHM} is a
formulation of the twistor transform for instantons on $\R^4.$ It
allows their construction in terms of linear algebra in a vector space
whose dimension is related to the instanton number. Since monopoles correspond
to infinite action instantons, then an adaptation of the ADHM construction
involving an infinite dimensional vector space, which can be represented
by functions of an auxiliary variable, might be possible. Nahm\cite{Na}
was able to formulate such an adaptation, which is now known as the 
Atiyah-Drinfeld-Hitchin-Manin-Nahm (ADHMN) construction, or the Nahm transform.

 The ADHMN construction is an equivalence between $k$-monopoles and Nahm data
$(T_1,T_2,T_3)$, which are three $k\times k$ matrices which depend
on a real parameter $s\in[0,2]$ and satisfy the following;\\

\newcounter{con}
\setcounter{con}{1}
(\roman{con})  Nahm's equation
\be
\frac{dT_i}{ds}=\half\epsilon_{ijk}[T_j,T_k], 
\label{nahms}
\ee\\

\addtocounter{con}{1}
(\roman{con}) $T_i(s)$ is regular for $s\in(0,2)$ and has simple
poles at $s=0$ and $s=2$,\\

\addtocounter{con}{1}
(\roman{con}) the matrix residues of $(T_1,T_2,T_3)$ at each
pole form the irreducible $k$-dimensional representation of SU(2),\\

\addtocounter{con}{1}
(\roman{con}) $T_i(s)=-T_i^\dagger(s)$,\\

\addtocounter{con}{1}
(\roman{con}) $T_i(s)=T_i^t(2-s)$.\\

Finding the Nahm data effectively solves the nonlinear part of the
monopole construction but in order to calculate the fields themselves
the linear part of the ADHMN construction must also be implemented\cite{Na,Hi2}. Given
Nahm data $(T_1,T_2,T_3)$ for a $k$-monopole we must solve the 
ordinary differential equation 
\be
({\identity}_{2k}\frac{d}{ds}+{\identity}_k\otimes \frac{x_j\sigma_j}{2}
+iT_j\otimes\sigma_j){\bf v}=0
\label{lin}
\ee
for the complex $2k$-vector ${\bf v}(s)$, where $\identity_k$ denotes
the $k\times k$ identity matrix and
${\bf x}=(x_1,x_2,x_3)$ is the point in space at which the monopole
fields are to be calculated. Introducing the inner product
\be
\langle{\bf v}_1,{\bf v}_2\rangle =\int_0^2 {\bf v}_1^\dagger{\bf v}_2\ ds
\label{ip}
\ee
then the solutions of (\ref{lin}) which are required are those which are
normalizable with respect to (\ref{ip}). It can be shown that the
space of normalizable solutions to (\ref{lin}) has (complex) dimension
two. If $\widehat {\bf v}_1,\widehat {\bf v}_2$ is an orthonormal basis
for this space then the Higgs field $\Phi$ and gauge potential $A_i$ are given by
\be
\Phi=i\left[ \begin{array}{cc}
\langle(s-1)\widehat {\bf v}_1,\widehat {\bf v}_1\rangle &
\langle(s-1)\widehat {\bf v}_1,\widehat {\bf v}_2\rangle \\
\langle(s-1)\widehat {\bf v}_2,\widehat {\bf v}_1\rangle &
\langle(s-1)\widehat {\bf v}_2,\widehat {\bf v}_2\rangle 
\end{array}
\right],
\
A_i=\left[ \begin{array}{cc}
\langle\widehat {\bf v}_1,\partial_i\widehat {\bf v}_1\rangle &
\langle\widehat {\bf v}_1,\partial_i\widehat {\bf v}_2\rangle \\
\langle\widehat {\bf v}_2,\partial_i\widehat {\bf v}_1\rangle &
\langle\widehat {\bf v}_2,\partial_i\widehat {\bf v}_2\rangle 
\end{array}
\right].
\label{higgs}
\ee

Nahm's equation (\ref{nahms}) has the Lax formulation
\be
\frac{d\Lambda}{ds}=[\Lambda,\Lambda_+]
\label{lax1}
\ee
where
\be
\Lambda=(T_1+iT_2)-2iT_3\zeta+(T_1-iT_2)\zeta^2, \ \
\Lambda_+=-iT_3+(T_1-iT_2)\zeta.
\label{lax2}
\ee
Hence the spectrum of  $\Lambda$ is $s$-independent, 
giving the associated algebraic curve 
\be
S=\mbox{det}(\eta+\Lambda)=0
\label{curve}
\ee
whose coefficients are the constants of motion associated with the
dynamical system (\ref{nahms}).
It can be shown\cite{Hi2} that this curve is again the same spectral curve 
that we have met twice already.

The power of the ADHMN construction may be demonstrated by considering the
case $k=1.$ In this case each Nahm matrix is just a purely imaginary function of
$s$, so the
solution of Nahm's equation (\ref{nahms}) is simply that each of these
functions must be constant. Thus the required Nahm data is simply $T_i=-ia_i/2$,
where the three real constants $a_i$ determine the position of the monopole  
to be ${\bf x}=(a_1,a_2,a_3).$ To construct a monopole at the origin the
Nahm data is thus $T_i=0$ and since this monopole is spherically symmetric 
we can simplify the presentation by restricting to the $x_3$-axis by setting
 ${\bf x}=(0,0,r).$

Writing ${\bf v}=(w_1,w_2)^t$ then the $2\times 2$ system (\ref{lin})
becomes the pair of decoupled equations
\be
\frac{dw_1}{ds}+\frac{r}{2}w_1=0, \ \ \frac{dw_2}{ds}-\frac{r}{2}w_2=0
\ee
which are elementary to solve as 
\be
w_1=c_1 e^{-rs/2}, \ \ w_2=c_2 e^{rs/2}
\ee
where $c_1$ and $c_2$ are arbitrary constants.

An orthonormal basis $\widehat {\bf v}_1,\widehat {\bf v}_2$, with respect
to the inner product (\ref{ip}), is obtained by the choice
\bea
c_1^2=0,& c_2^2=r/(e^{2r}-1) & \mbox{for} \ \widehat {\bf v}_1\\
c_2^2=0,& c_1^2=r/(1-e^{-2r}) & \mbox{for} \ \widehat {\bf v}_2.
\eea
Note that a different choice of orthonormal basis corresponds to a different
choice of gauge for the monopole fields. In the gauge we have chosen
$\Phi=i\varphi \sigma_3$ where
\be
\varphi=\langle(s-1)\widehat {\bf v}_1,\widehat {\bf v}_1\rangle
=\frac{r}{(e^{2r}-1)}\int_0^2 (s-1)e^{rs} \ ds 
=\frac{r\mbox{cosh} r-\mbox{sinh} r }{r \mbox{sinh} r }
\ee
which reproduces the expression (\ref{k1}).

After a suitable orientation the Nahm data of a 2-monopole has the form
\be
 T_1=\frac{f_1}{2}\left(\begin{array}{cc} 0& i \\
i&0\end{array}\right),\; T_2=\frac{f_2}{2}\left(\begin{array}{cc} 0& 1 \\
-1&0\end{array}\right),\; T_3=\frac{f_3}{2}\left(\begin{array}{cc} -i& 0 \\
0&i\end{array}\right) \label{nd2}\ee
with corresponding spectral curve
\be
{ \eta}^{2} + \frac{1}{4}\left( 
(f_1^2-f_2^2) + (2f_1^2+2f_2^2 - 4f_3^2) \zeta^2+ 
(f_1^2-f_2^2) \zeta^4\right)=0.
\label{curve1}
\ee
With this ansatz Nahm's equation reduces to the Euler top equation
\be
\frac{df_1}{ds}=f_2f_3
\ee
and cyclic permutations. The solution satisfying the appropriate
Nahm data boundary conditions is\cite{BPP}
\be
 f_1=-\frac{K\mbox{dn}(Ks)}{\mbox{sn}(Ks)},\;\;
     f_2=-\frac{K}{\mbox{sn}(Ks)},\;\;
     f_3=-\frac{K\mbox{cn}(Ks)}{\mbox{sn}(Ks)}
\label{ndk2}
\ee
where $\mbox{sn}(u),\mbox{cn}(u),\mbox{dn}(u)$ denote the Jacobi elliptic
 functions with argument $u$ and parameter $m$,
and, as earlier, $K$ is the complete elliptic integral of the first kind with
parameter $m.$
Substituting these expressions for $f_i$ into the curve (\ref{curve1})
and using the standard identities
\hbox{$\mbox{sn}^2(u)+\mbox{cn}^2(u)=1$}
and \hbox{$m\mbox{sn}^2(u)+\mbox{dn}^2(u)=1$}
 we once again obtain the 2-monopole spectral curve (\ref{sc2b}).

The fact that the spectral curve of a $k$-monopole has genus $(k-1)^2$
means that Nahm's equation, whose flow is linearized on the Jacobian of this
curve, can be solved in terms of theta functions defined on a Riemann surface
of genus $(k-1)^2.$ For $k=2$ then the curve is elliptic, which explains why
the general solution of Nahm's equation can be obtained in terms of elliptic
functions and why the parameters in the spectral curve are obtained in
terms of elliptic integrals. However, for $k>2$ it appears a very
difficult task to attempt to express the general solution of Nahm's equation
in terms of theta functions and then try and impose the required boundary
conditions necessary to produce Nahm data. This is the underlying mathematical
obstruction which has prevented the general 3-monopole solution from being
constructed, or even its spectral curve.

Nonetheless, there are simplifying special cases where extra symmetry of the monopole
means that progress can be made. In such symmetric cases, where the monopole has
some rotational symmetry given by a group $G\subset SO(3)$, then it is
not the genus of the spectral curve $S$ which is the important quantity, but
rather the genus $\tilde g$ of the quotient curve $\tilde S=S/G.$ 
If $\tilde g<2$ then the situation is greatly simplified and there is a hope
of some form of construction (either the monopole fields, Nahm data or spectral
curve) in terms of, at worst, elliptic functions and integrals. 
The axially symmetric $k$-monopoles for $k>2$, given by
 (\ref{scaxeven}) and (\ref{scaxodd}), are such
examples and we shall see some other recent examples in Section 3, where $G$ is
a Platonic symmetry group.

By making use of the Nahm transform, Donaldson\cite{Do} was able to prove
that the monopole moduli space ${\cal M}_k$ is diffeomorphic to the
space of degree $k$ based rational maps \hbox{$R:$ \CP$^1\mapsto$ \CP$^1.$}
Explicitly, Donaldson showed how every $k$-monopole gives rise to
a unique rational map 
\be
R(z)=\frac{p(z)}{q(z)} \ee
where $q(z)$ is a
monic polynomial of degree $k$ in the complex variable $z$ and 
$p(z)$ is a polynomial of degree less than $k$, with no factors
in common with $q(z)$. 

To understand this diffeomorphism better it is useful to follow the analysis
of Hurtubise\cite{Hu2}. We have already noted that a study of Hitchin's equation
(\ref{hitchins}) along a line shows that there are two independent solutions
with a basis $(v_0,v_1)$ consisting of solutions which are respectively
bounded and unbounded as $u$, the coordinate along the line, tends to infinity.
In addition we introduced the basis $(v_0',v_1')$ of bounded and unbounded
solutions in the opposite direction $u\rightarrow -\infty.$
Now fix a direction in $\R^3$,
which gives the decomposition
\be
\R^3\cong\C\times\R.
\label{decomp}
\ee
For convenience, we choose this direction to be that of the positive $x_3$-axis
 and denote by $z$ the
complex coordinate on the $x_1x_2$-plane. Thus the coordinate $u$ in the above
analysis of Hitchin's equation (\ref{hitchins}) is now $x_3.$
The approach of Hurtubise is to consider the
scattering along all such lines and write
\begin{eqnarray}
 v_0^{\prime}&=&a(z)v_0+b(z)v_1,\\
  v_0&=&a^{\prime}(z)v_0^{\prime}+b(z)v_1^{\prime}.
\end{eqnarray}
The rational map is then given by
\be
 R(z)=\frac{a(z)}{b(z)}.
\ee
Furthermore, since the spectral curve $S(\eta,\zeta)$ of a monopole corresponds
to the bounded solutions of (\ref{hitchins}) then
\be b(z)=S(z,0).\ee
Finally, it can be shown\cite{AH} that the full
scattering data is given by
\be 
\left[\begin{array}{cc} a & b\\-b^\prime &
-a^{\prime}\end{array}\right]\left(\begin{array}{c}v_0\\v_1
\end{array}\right)\ =\left(\begin{array}{c}v_0^{\prime}\\v_1^{\prime}
\end{array}\right)
\ee
where 
\be 
aa^{\prime}=1+b^\prime b.
\ee

The advantage of rational maps is that monopoles
are easily described in this approach, since one simply
writes down any rational map. The disadvantage
is that the rational map tells us very little about the monopole. In
particular, since
the construction of the rational map requires the choice of a
direction in $\R^3$ it is not  possible to study the full symmetries
of a monopole from its rational map. However, the following isometries are
known\cite{HMM}. Let $\lambda\in U(1)$ and $\nu \in \C$ define a
rotation and translation respectively in the plane $\C$. Let $x\in\R$
define a translation perpendicular to the plane and let $\mu\in U(1)$
be a constant gauge transformation. Under the composition of these
transformation a rational map $R(z)$ transforms as
\be R(z)\mapsto \mu^2e^{2x}\lambda^{-2k}R(\lambda^{-1}(z-\nu)).\label{trans1}\ee
Furthermore, under the reflection
$x_3\mapsto -x_3$, 
 $R(z)=p(z)/q(z)$ transforms as
\be \frac{p(z)}{q(z)}\mapsto\frac{I(p)(z)}{q(z)}\label{trans2}\ee
where $I(p)(z)$ is the unique polynomial of degree less than $k$ such
that $(I(p)p)(z)=1$ mod $q(z)$. 

Some information regarding the monopole configuration can be determined from
the rational map in special cases, corresponding to well-separated monopoles.

Bielawski\cite{Bi1} has proved that for a rational map $p(z)/q(z)$ with
 well-separated poles $\beta_1,\ldots,\beta_k$ the corresponding monopole is
approximately composed of unit charge monopoles located at the points
$(x_1,x_2,x_3)$, where $x_1+ix_2=\beta_i$ and 
$x_3=\frac{1}{2}\log{|p(\beta_i)|}$.
This approximation applies only when
the values of the numerator at the poles is small compared to the
distance between the poles. 

For the complimentary case of monopoles strung
 out in well-separated clusters along (or nearly along) the $x_3$-axis, the 
large $z$ expansion of the rational map $R(z)$ is\cite{AH,HS2}
\be 
R(z)\sim\frac{e^{2x+i\chi}}{z^L}+\frac{e^{2y+i\phi}}{z^{2L+M}}+...
\label{paulsform}
\ee
where $L$ is the
charge of the topmost cluster with $x$ its elevation above the
plane and $M$ is the charge of the next highest cluster with elevation $y.$

On a technical point, it is often
convenient to restrict to what are known as strongly centred monopoles\cite{HMM},
which roughly means that the total phase is unity and the centre of mass
is at the origin. More precisely, in terms of the rational map a monopole is
 strongly centred if and only if the roots of $q$ sum to zero and the product
 of $p$ evaluated at each of the roots of $q$ is equal to one.

Although this rational map description of monopoles is very useful, particularly
as a convenient description of the monopole moduli space, it does suffer from
the drawback of requiring a choice of direction in $\R^3$, which is not helpful
in several contexts. Atiyah has suggested that a second rational map description
might exist, in which the full rotational symmetries around the origin of $\R^3$
 are not broken,
and indeed this has recently been confirmed by Jarvis\cite{Ja}. To construct
the Jarvis rational map one considers Hitchin's equation (\ref{hitchins})
along each radial line from the origin to infinity. The coordinate $u$ is now
identified with the radius $r$ and as we have already seen there is just one
solution, with basis $v_0$, which is bounded as $r\rightarrow\infty.$ Writing
$v_0(r)=(w_1(r),w_2(r))^t$ then define $J$ to be the ratio of these
components at the origin ie. $J=w_1(0)/w_2(0).$ Now take $z$ to determine
a particular radial line, by giving its direction as a point on the Riemann sphere.
Then, in fact, $J$ is a holomorphic function of $z$, since by virtue of
the Bogomolny equation (\ref{bog}) the covariant derivative in the angular
direction, $D_{\bar z}$, commutes with the operator $D_r-i\Phi$ appearing
in Hitchin's equation (\ref{hitchins}). It can be verified that $J$ is a rational
function of degree $k$ so that we have a new rational map 
\hbox{$J:$ \CP$^1\mapsto$ \CP$^1.$} Note that the Jarvis rational map is unbased,
since a gauge transformation replaces $J$ by an $SU(2)$ M\"obius transformation
determined by the gauge transformation evaluated at the origin. Thus the 
correspondence is between a monopole
and an equivalence class of Jarvis maps, where two maps are equivalent 
if they can be mapped into each other by a reorientation of the target Riemann
sphere. In Section 3 we shall see some explicit examples of Jarvis maps
corresponding to symmetric monopoles.

\news
\vspace*{1pt}\textlineskip	
\section{Platonic Monopoles}	
\vspace*{-0.5pt}
\noindent
In the last Section we commented on how the construction of monopoles can
be simplified by restricting to special symmetric cases. In this Section
we discuss in detail some particular examples of this, where the symmetry
group $G$ is Platonic ie. tetrahedral, octahedral or icosahedral.

The first issue to confront is how to impose a given symmetry 
in the monopole construction and this is most easily discussed in terms
of the spectral curve approach. Recall that in terms of the coordinates 
$(\eta,\zeta)$ on \mt\ the lines through the origin are parameterized by
$\zeta$ with $\eta=0.$ An $SO(3)$ rotation in $\R^3$ acts on the Riemann
sphere coordinate via an $SU(2)$ M\"obius transformation and since
$\eta$ is the coordinate in the tangent space to this Riemann sphere then
it transforms via the derivative of this M\"obius transformation.
For example,
rotation through an angle  $\phi$ around the $x_3$-axis is given by
\be
 R_\phi:(\eta,\zeta)\mapsto (\eta',\zeta')=(e^{i\phi}\eta,e^{i\phi}\zeta).
\label{rotx3}
\ee
A monopole is symmetric under a symmetry group $G$ if its transformed
spectral curve $S(\eta',\zeta')=0$ is the same as its original curve
$S(\eta,\zeta)=0$, for all M\"obius transformations corresponding to the
elements of $G.$ For example, it is now clear that the spectral curves
(\ref{scaxeven}) and (\ref{scaxodd}) describe monopoles which are axially
symmetric, since they are invariant under the transformation (\ref{rotx3})
for all $\phi$, due to the fact that these curves are homogeneous in the
twistor coordinates.

It is thus straightforward to write down, for example, that the candidate
axially symmetric 2-monopole spectral curve has the form
\be
\eta^2+a\zeta^2=0
\ee
where $a$ is real due to the reality constraint (\ref{reality}).
However, symmetry alone gives no information regarding the possible
values (if any) of $a$. This is determined by the non-singularity 
condition which can be analyzed directly\cite{Hu1} or, more easily, using
the ADHMN formulation as detailed in the previous Section. Either way, the
result that $a=\pi^2/4$ is obtained, which also proves the non-existence
of a spherically symmetric 2-monopole, which would require $a=0$, to be
invariant under all $SU(2)$ M\"obius transformations.

It is certainly not obvious what symmetry to attempt to impose in order
to simplify the  monopole construction, but the proposal by Hitchin, Manton
\& Murray\cite{HMM} to consider Platonic symmetries turns out to be a very
fruitful one. Its motivation lies with some numerical work of Braaten, Townsend
and Carson\cite{BTC} on another kind of topological soliton, the Skyrmion, where
it was found that Skyrmions of charge three and four appear to have tetrahedral and
cubic symmetry respectively. 

Let us first consider the tetrahedral case, $G=T_d.$ One way to proceed would
be as described above, that is, to write down the generators of $T_d$, compute
the associated $SU(2)$ M\"obius transformations and hence derive the
form of the invariant polynomials which are allowed in the candidate spectral
curve. However, there is a short cut available, since the result which one
would obtain by this method is precisely the computation of the invariant tetrahedral
Klein polynomial\cite{Kl}. The Klein polynomials associated with a Platonic
solid are the polynomials obtained by taking either the vertices, faces or
edges of the solid, projecting these onto the unit 2-sphere and computing
the monic polynomial whose complex roots are exactly these points, when thought
of as points on the Riemann sphere. These Klein polynomials are all listed
in ref.\cite{Kl} and of relevance to the tetrahedral case is 
\be
{\cal T}_e=\zeta(\zeta^4-1)
\label{kpte}
\ee
which is the edge polynomial of a (suitably oriented) tetrahedron.
Note that this polynomial should really be thought of projectively as a degree six
polynomial with one root at infinity.
The other two tetrahedral Klein polynomials
\be
{\cal T}_{v,f}=\zeta^4\pm2\sqrt{3}i\zeta^2+1
\label{kptvf}
\ee
corresponding to vertex and face points, are not appropriate in this case
since they are not strictly invariant, as they pick up a multiplying factor
under some of the tetrahedral transformations.

From the restriction that the coefficients $a_r(\zeta)$ in the general curve
(\ref{gencurve}) must be a polynomial of maximum degree $2r$, then it is clear
that in order to be able to accommodate the tetrahedral term (\ref{kpte}) requires
$k\ge 3.$ Hence the smallest charge candidate tetrahedral spectral curve is
the charge three curve
\be
\eta^3+ic_3{\cal T}_e=0
\ee
where $c_3$ is a real constant to satisfy reality. 
The determination of the possible (if any) values of $c_3$ that are allowed
by non-singularity is a more difficult problem, to which we shall return shortly,
but the result is that\cite{HMM}
\be
 c_3=\pm \frac{\Gamma(1/6)^3\Gamma(1/3)^3}{48\sqrt{3}\pi^{3/2}}
\ee
where the $\pm$ corresponds to the tetrahedron and its dual. This proves the
existence and uniqueness, up to translations and rotations, of a tetrahedrally
symmetric 3-monopole. We shall study this monopole and other symmetric
examples in more detail at the end of this Section.

Turning now to the octahedral group $G=O_h$, the relevant Klein polynomial
is the face polynomial of the octahedron
\be
{\cal O}_f=\zeta^8+14\zeta^4+1.
\ee
Since this is a degree eight polynomial then the charge is required to satisfy $k\ge 4$,
with the smallest charge case having the form
\be
\eta^4+c_4{\cal O}_f=0
\ee
for $c_4$ real. Again a non-singularity investigation\cite{HMM} reveals the unique
value\footnote{In refs.\cite{HMM,HS2} there is a factor of 16 error.} 
\be
c_4=\frac{3\Gamma(1/4)^8}{1024\pi^2}.
\ee
There is also a unique octahedrally symmetric 5-monopole\cite{HS2} given by$^a$
\be
\eta^5+4c_4\eta{\cal O}_f=0.
\ee
In the icosahedral case, $G=I_h$, all the Klein polynomials are strictly invariant,
thus the smallest degree polynomial to use is the icosahedron vertex polynomial
\be
{\cal I}_v=\zeta(\zeta^{10}+11\zeta^5-1).
\label{kpiv}
\ee
This implies that $k\ge 6$ so the smallest charge candidate is
\be
\eta^6+c_6{\cal I}_v=0.
\ee
However, it is rather surprising to find\cite{HMM} that there are no
allowed values of $c_6$ compatible with non-singularity and hence no
icosahedrally symmetric monopoles exist with $k\le 6.$

A unique icosahedrally symmetric 7-monopole does exist\cite{HS2} with
\be
\eta^7+c_7\eta{\cal I}_v=0,\ \mbox{where} \
c_7=\frac{\Gamma(1/6)^6\Gamma(1/3)^6}{64\pi^3}.
\label{dodec}
\ee

For the case of the axially symmetric 2-monopole we have seen that the
simplest method of determining the value taken by the constant in the spectral curve
is through the computation of the Nahm data, from which the 
 spectral curve can be read off using (\ref{curve}). By a simple application
of the Riemann-Hurwitz formula it can be shown\cite{HMM} that in all the Platonic
examples given above the quotient curve $\tilde S=S/G$ is elliptic. Thus the
corresponding reduction of Nahm's equation is solvable in terms of elliptic
functions and this provides the easiest method of computing the spectral
curve coefficients. 

With this aim in mind we need to formulate an algorithm for constructing
Nahm data invariant under any discrete symmetry group $G\subset SO(3).$
This is explained in\cite{HMM,HS2} and we review it below.

The Nahm matrices are
traceless, so they transform under the rotation group as
\begin{eqnarray}\underline{3}\otimes 
sl(\underline{k}) &\cong&\underline{3}\otimes
 (\underline{2k-1}\oplus\underline{2k-3}\oplus
 \ldots \oplus \underline{3})\nonumber\\
&\cong&( \underline{2k+1}_u\oplus \underline{2k-1}_m\oplus
\underline{2k-3}_l)\oplus \ldots\nonumber\\ 
&\ldots& \oplus ( \underline{2r+1}_u\oplus \underline{2r-1}_m\oplus
\underline{2r-3}_l)\oplus\ldots\oplus (
\underline{5}_u\oplus \underline{3}_m\oplus\underline{1}_l)\nonumber\\
\label{decomp2}
\end{eqnarray}
where $\underline{r}$ denotes the unique irreducible $r$-dimensional
representation of $su(2)$ and the subscripts $u,m$ and $l$ 
(which stand for upper, middle and lower) are a
convenient notation 
allowing us to distinguish between $(2r+1)$-dimensional
representations occurring as 
\begin{eqnarray*}\underline{3}\otimes\underline{2r-1}&\cong&\underline{2r+1}_u\oplus \underline{2r-1}_m\oplus
\underline{2r-3}_l,\\ \underline{3}\otimes\underline{2r+1}&\cong&\underline{2r+3}_u\oplus \underline{2r+1}_m\oplus
\underline{2r-1}_l\end{eqnarray*}
and 
$$\underline{3}\otimes\underline{2r+3}\,\,\,\cong\,\,\,\underline{2r+5}_u\oplus \underline{2r+3}_m\oplus
\underline{2r+1}_l.$$

We can then use invariant homogeneous polynomials over \CP$^1$, that is we
use the  homogeneous coordinates $\zeta_1/\zeta_0=\zeta$, to
construct $G$-invariant Nahm triplets.

 The vector space of degree $2r$ homogeneous polynomials
$a_{2r}\zeta_1^{2r}+a_{2r-1}\zeta_1^{2r-1}\zeta_0+\ldots+a_0\zeta_0^{2r}$ is the carrier space for $\underline{2r+1}$
under the identification
\be X=\zeta_1 \frac{\partial}{\partial \zeta_0},\qquad Y=\zeta_0
\frac{\partial}{\partial \zeta_1},\qquad H=-\zeta_0
\frac{\partial}{\partial \zeta_0}+\zeta_1 \frac{\partial}{\partial
\zeta_1}\ee
where $X,Y$ and $H$ are the basis of $su(2)$ satisfying
\be [X,Y]=H,\qquad[H,X]=2X,\qquad[H,Y]=-2Y.\label{HXY}\ee
If $p(\zeta_0,\zeta_1)$ is a $G$-invariant
homogeneous polynomial we can construct a $G$-invariant
$\underline{2r+1}_u$ charge $k$ Nahm triplet by the following scheme. 

\newcounter{sch}
\setcounter{sch}{1}
(\roman{sch}) The inclusion 
\be\underline{2r+1}\hookrightarrow\underline{3}\otimes
  \underline{2r-1}\cong\underline{2r+1}_u\oplus
\underline{2r-1}_m\oplus\underline{2r-3}_l\ee
is given on polynomials by
\be p(\zeta_0,\zeta_1)
\mapsto\xi_1^2\otimes p_{11}(\zeta_0,\zeta_1)+2\xi_0 \xi_1 \otimes p_{10}
(\zeta_0,\zeta_1)+\xi_0^2\otimes 
p_{00}(\zeta_0,\zeta_1)\ee
where we have used the notation
\be p_{ab}(\zeta_0,\zeta_1)=\frac{\partial^2 p}{\partial\zeta_a\partial\zeta_b}(\zeta_0,\zeta_1).\ee

\addtocounter{sch}{1}
(\roman{sch}) The polynomial expression $\xi_1^2\otimes
p_{11}(\zeta_0,\zeta_1)+2\xi_0\xi_1\otimes p_{10}(\zeta_0,\zeta_1)+\xi_0^2\otimes 
p_{00}(\zeta_0,\zeta_1)$ is rewritten in the form
\be\xi_1^2\otimes
q_{11}(\zeta_0\frac{\partial}{\partial\zeta_1})\zeta_1^{2r}
+(\xi_0\frac{\partial}{\partial
  \xi_1})\xi_1^2\otimes
q_{10}(\zeta_0\frac{\partial}{\partial\zeta_1})\zeta_1^{2r}
+\half (\xi_0\frac{\partial}{\partial
  \xi_1})^2\xi_1^2\otimes 
q_{00}(\zeta_0\frac{\partial}{\partial\zeta_1})\zeta_1^{2r}.\ee

\addtocounter{sch}{1}
(\roman{sch}) This then defines a triplet of $k\times k$ matrices. Given a
$k\times k$ representation of $X,Y$ and $H$ above, the invariant Nahm
triplet is given by:
\be (S_1^\prime,S_2^\prime,S_3^\prime)=
(q_{11}(\mbox{ad}Y)X^{r},q_{10}(\mbox{ad}Y)X^{r},q_{00}(\mbox{ad}Y)X^{r}),
\ee
where ad$Y$ denotes the adjoint action of $Y$ and is given on a
general matrix $M$ by ad$YM=[M,Y]$.

\addtocounter{sch}{1}
(\roman{sch}) The Nahm isospace basis is transformed. This
transformation is given by 
\be(S_1,S_2,S_3)=(\half S_1^\prime+S_3^\prime,-\frac{i}{2}
S_1^\prime+iS_3^\prime,-iS_2^\prime).\label{basis}\ee
Relative to this basis the $SO(3)$-invariant Nahm triplet
corresponding to the $\underline{1}_l$ representation in (\ref{decomp2})
is given by $(\rho_1,\rho_2,\rho_3)$ where 
\be
\rho_1=X-Y, \qquad \rho_2=i(X+Y), \qquad \rho_3=iH.
\ee

It is also necessary to construct invariant Nahm triplets lying in
the $\underline{2r+1}_m$ representations. To do this, we first
construct the corresponding $\underline{2r+1}_u$ triplet. We then
write this triplet in the canonical form
\begin{eqnarray} [c_0+c_1(\mbox{ad}Y\otimes 1+1\otimes \mbox{ad} Y
  )&+&
\ldots
+c_i(\mbox{ad}Y\otimes 1+1\otimes \mbox{ad}
Y)^i \\ &+&\ldots+c_{2r}(\mbox{ad}Y\otimes 1+1\otimes \mbox{ad} Y
)^{2r}]\,X\otimes X^r\nonumber\label{canon}\end{eqnarray} 
and map this isomorphically into $\underline{2r+1}_m$ by mapping
the highest weight vector $X\otimes X^r$ to the highest weight vector
\be X\otimes \mbox{ad}YX^{r+1}-\frac{1}{r+1}\mbox{ad}YX\otimes X^{r+1}.\ee

As an example, applying this scheme to determine charge seven Nahm data
with icosahedral symmetry, we take the icosahedron vertex polynomial ${\cal I}_v$
(\ref{kpiv}) in homogeneous form 
\be
{\cal I}_v=\zeta_1^{11}\zeta_0+11\zeta_1^6\zeta_0^6-\zeta_1\zeta_0^{11}.
\ee
This leads to an icosahedrally invariant Nahm triplet
\be
T_i(s)=x(s)\rho_i+z(s)Z_i
\label{ndred}
\ee
where 
\begin{eqnarray}
Z_1\hskip -0.2cm & =\hskip -0.2cm & \:\; \left[ 
\begin{array}{ccccccc}
0          &5\sqrt {6} & 0              & 0            &
7\sqrt{6}\sqrt {10} & 0 & 0  \\
-5\sqrt{6}& 0          &- 9\sqrt {10}  & 0            & 0                    & 0 & 0 \\
0          &9\sqrt {10}& 0              &5\sqrt {12}  & 0    & 0 &  -7\sqrt {6}\sqrt {10} \\
0 & 0 & -5\sqrt{12}& 0& 5\sqrt {12}& 
0& 0 \\
 - 7\sqrt{6}\sqrt {10}& 0& 0&  - 5\sqrt {12}& 
0&  - 9\sqrt {10}&0 \\
0& 0& 0& 0&9\sqrt {10}& 0& 5
\sqrt {6} \\
0&0&7\sqrt {6}\sqrt {10}& 0&0&  - 
5\sqrt {6}&0
\end{array}
 \right] \nonumber \\
 Z_2\hskip -0.2cm & =\hskip -0.2cm & i  \left[ 
\begin{array}{ccccccc}
0&5\sqrt {6}&0&0&-7\sqrt {6}\sqrt {10}&0&0\\
5\sqrt{6}&0&-9\sqrt{10}&0
&0&0&0 \\
0&-9\sqrt{10}&0&5\sqrt{12}
&0&0&7\sqrt {6}\sqrt {10} \\
0&0&5\sqrt {12}&0&5\sqrt {
12}&0&0\\
 -7\sqrt {6}\sqrt {10}& 0& 0& 5\sqrt {
12}& 0&  - 9\sqrt {10}& 0 \\
0&0&0&0&-9\sqrt {10}&0&
5\sqrt {6} \\
0&0&7\sqrt{6} \sqrt {10}&0&0&5\sqrt {6}&0
\end{array}
 \right] \nonumber \\
 Z_3\hskip -0.2cm & =\hskip -0.2cm & i \left[ 
\begin{array}{ccccccc}
 - 12&0&0&0&-14\sqrt {
6}&0&0 \\
0&48&0&0&0&0&- 14\sqrt {6} \\
0&0&-60&0&0&0&0 \\
0&0&0&0&0&0&0 \\
0&0&0&0&60&0&0 \\
 - 14\sqrt {6}&0&0&0&0&-48& 0
 \\
0&- 14\sqrt {6}&0&0&0&0&12
\end{array}
 \right]
\end{eqnarray}
and the basis has been chosen in which
\be
 H=\mbox{diag}(6,4,2,0,-2,-4,-6)
\ee
\be
X=Y^t=  \left[ 
{\begin{array}{rcccccc}
0 & \sqrt {6} & 0 & 0 & 0 & 0 & 0 \\
0 & 0 & \sqrt {10} & 0 & 0 & 0 & 0 \\
0 & 0 & 0 & \sqrt {12} & 0 & 0 & 0 \\
0 & 0 & 0 & 0 & \sqrt {12} & 0 & 0 \\
0 & 0 & 0 & 0 & 0 & \sqrt {10} & 0 \\
0 & 0 & 0 & 0 & 0 & 0 & \sqrt {6} \\
0 & 0 & 0 & 0 & 0 & 0 & 0
\end{array}}
 \right].
\ee
Substitution of the Nahm data (\ref{ndred}) into Nahm's equation (\ref{nahms})
produces the reduced equations
\begin{eqnarray}
\frac{d x}{ds}&=&2x^2-750z^2\label{n1i}\\
  \frac{dz}{d s}&=&-10xz+90z^2\nonumber
\end{eqnarray}
with corresponding spectral curve
\be
 \eta^7+c_7\eta{\cal I}_v=0
\ee
where
\be
 c_7=552960(14xz-175z^2)(x+5z)^4
\ee
is a constant.

These equations have the solution 
\begin{eqnarray}
x(s)&=&\frac{2\kappa}{7}\left[-3\sqrt{\wp(2\kappa s)}+
\frac{{\wpp}(2\kappa s)}{4\wp(2\kappa s)}\right]\\
z(s)&=&-\frac{\kappa}{35}\left[\sqrt{\wp(2\kappa s)}+
\frac{{\wpp}(2\kappa s)}{2\wp(2\kappa s)}\right]
\end{eqnarray}
where $\wp(t)$ is the Weierstrass function satisfying 
\be
{\wpp}^2=4(\wp^3-1)
\ee
and to satisfy the Nahm data boundary conditions $\kappa$ is the real half-period
of this elliptic function ie.
\be
\kappa=\frac{\Gamma(1/6)\Gamma(1/3)}{8\sqrt{3\pi\,}}.
\ee
Then 
\be
 c_7=110592\kappa^6=\frac{\Gamma(1/6)^6\Gamma(1/3)^6}{64\pi^3}
\ee
and we obtain the icosahedrally symmetric 7-monopole
spectral curve (\ref{dodec}).

In a similar way the Nahm data and spectral curves of the other Platonic
monopoles given earlier can be computed. Once the Nahm data is known it would
be nice to complete the linear part of the ADHMN construction and derive
the monopole gauge fields explicitly. However, the form of the Nahm data
in these examples is sufficiently complicated that the explicit solution
of equation (\ref{lin}) appears a difficult task. Fortunately, a numerical solution
of this equation and a subsequent numerical implementation of the linear part
of the ADHMN construction can be achieved\cite{HS1}, to display surfaces
of constant energy density.

In Fig.1 we display surfaces of constant energy density (not to scale) 
for the four Platonic monopoles discussed in this Section.

\begin{figure}[htbp]
\epsfxsize=12cm \epsffile{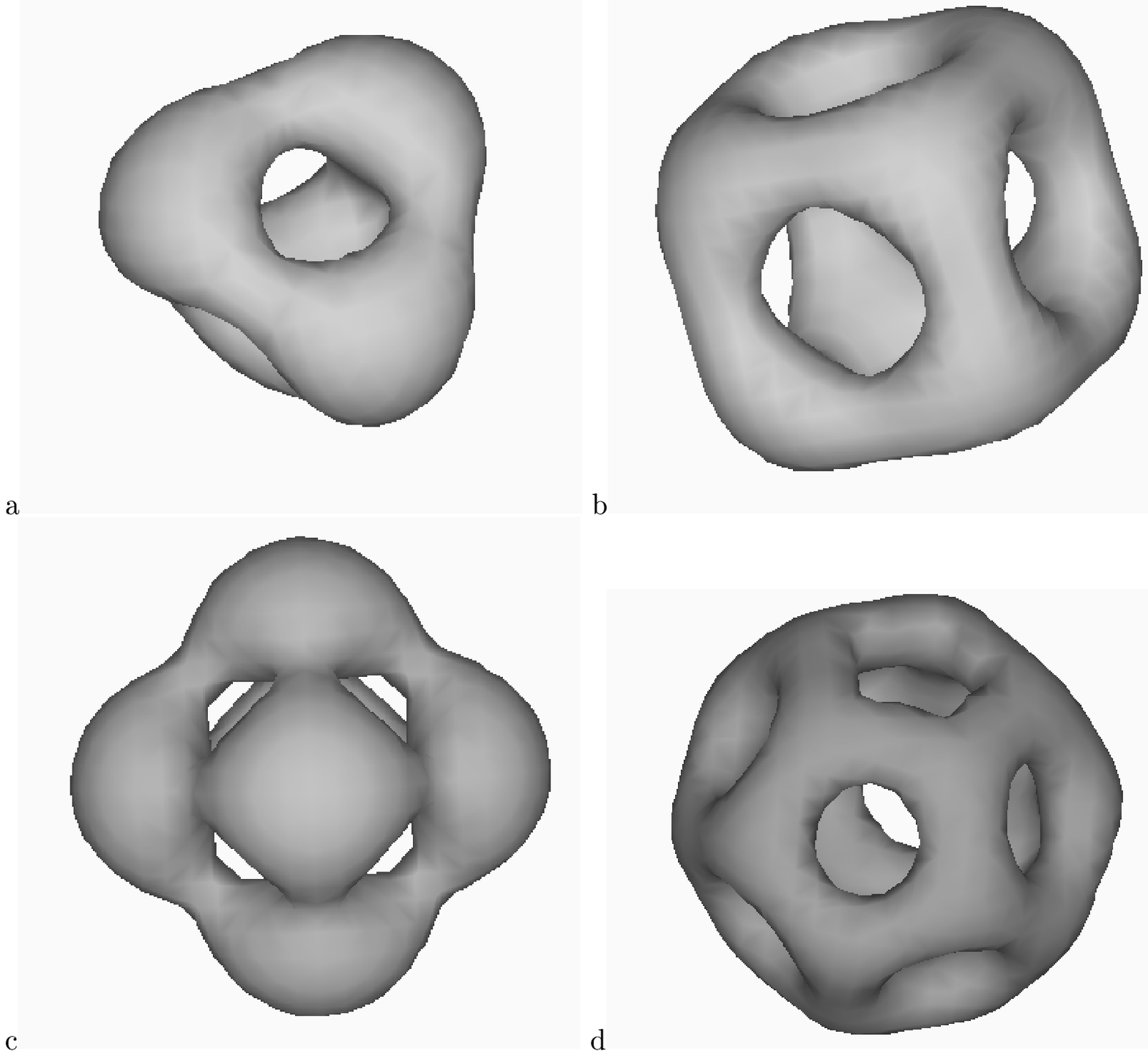}
\ \vskip -3cm
\fcaption{Energy density surfaces for
a) Tetrahedral 3-monopole; b) Cubic 4-monopole; c) Octahedral 5-monopole; 
d) Dodecahedral 7-monopole. }
\end{figure}

We see that each of the monopoles resembles a Platonic solid (tetrahedron, cube,
octahedron and dodecahedron for $k=3,4,5,7$ respectively) with the energy
density taking its maximum value on the vertices of this solid.

It is interesting to ask where the zeros of the Higgs field occur. 
We know from the general discussion in Section 1 that the number of Higgs
zeros, when counted with multiplicity, is $k.$ Consider then, for example, the
tetrahedral 3-monopole.  It is clear that the only way to
arrange three points with tetrahedral symmetry is to
put all three points at the origin. Thus if the 
tetrahedral monopole has three zeros of the Higgs field
then they must all be at the origin, as in the case
of the axisymmetric 3-monopole. This would be a little intriguing, but the
true situation is even more interesting. In fact, by examining the Higgs
field from the numerical ADHMN construction, it transpires that there
are five zeros of the Higgs field\cite{HS3}. 
 There are four positive zeros (ie. each corresponding to a local winding of $+1$)
on the vertices of a regular tetrahedron and an
anti-zero (ie. corresponding to a local winding of $-1$) at the
origin. Here a local winding at a point is defined as the winding
number of the normalized Higgs field, $\Phi/\|\Phi\|$, on a
small 2-sphere centred at this point.
This integer winding number counts the number of zeros of the Higgs
field, counted with multiplicity, inside this 2-sphere and thus, 
 by definition, the sum of these local windings around all Higgs zeros
must equal $k.$
Therefore the tetrahedral 3-monopole is
 a solution in which the Higgs field has both positive multiplicity and 
negative multiplicity zeros but nonetheless saturates the Bogomolny
energy bound. This is a recently discovered new phenomenon for monopoles
which is still not fully understood.

One possible approach to investigating this interesting phenomenon
may be to examine analogous cases for monopoles in  hyperbolic spaces\cite{At2}.
Just as monopoles in $\R^3$ may be interpreted as self-dual gauge
fields in $\R^4$ with a translation invariance, 
then hyperbolic monopoles may be identified as self-dual gauge fields which
are invariant under a circle action. To be  topologically correct one should really
consider the self-dual gauge fields to be defined in the compactification
of $\R^4$ to $S^4$, then a circle action
leaves invariant an $S^2.$ A subclass of self-dual gauge fields can be obtained
from the Jackiw-Nohl-Rebbi ansatz\cite{JNR}, which, to determine an $n$-instanton,
requires a choice of $n+1$ points in $\R^4$ together with $n+1$ weights.
By choosing equal weights and placing the points suitably in the invariant $S^2$,
we could construct symmetric hyperbolic monopoles. For example, taking $n=3$
and placing the four points on the vertices of a regular tetrahedron in the 
invariant $S^2$ a tetrahedrally invariant hyperbolic monopole can be computed.
The advantage of working with hyperbolic monopoles is that the  Jackiw-Nohl-Rebbi
instanton has a simple and explicit form, so that it should be possible to
derive an explicit expression for the hyperbolic monopole fields. With
such an explicit expression it may be possible to understand the monopole
fields better and hopefully learn something about the Euclidean case.

An examination of the other Platonic monopoles reveals\cite{Su5} that in
some cases these extra anti-zeros are present and in others they are not.
In Section 4, when we investigate monopole dynamics, we shall discuss some
aspects of these extra Higgs zeros.

After the existence of these Platonic monopoles was proved in the above
way, using the Nahm transform, the new rational map correspondence of Jarvis, 
which we discussed in Section 2, was proved. This allows a much easier
study of the existence of monopoles with rotational symmetries, since there
are no differential equations which need to be solved. Recall that the
Donaldson rational map is of no use in tackling this problem, since it
requires the choice of a direction in $\R^3$, thereby breaking the
rotational symmetry of the problem.

A Jarvis map, \hbox{$J:$ \CP$^1\mapsto$ \CP$^1$,} is symmetric under a subgroup 
$G\subset SO(3)$ if there is a set of M\"obius transformation pairs $\{g,D_g\}$ 
with $g\in G$ acting on the domain and $D_g$ acting on the target, such that
\be
J(g(z))=D_gJ(z).
\label{inv}
\ee
where the transformations $\{D_g\}$ form a 2-dimensional representation of $G.$

The simplest case is the spherically symmetric 1-monopole, given by $J=z$. 
The axially symmetric $k$-monopole, with symmetry around the $x_3$-axis, has the Jarvis
map $J=z^k$, for $k\ge 2.$

As a more complicated example, let us construct the Jarvis map of a tetrahedrally symmetric
3-monopole and hence prove its existence. 

To begin with, we can impose $180^\circ$ rotation symmetry about all three
Cartesian axes by requiring the symmetries under
$z\mapsto -z$ and $z\mapsto 1/z$ as
\be
J(-z)=-J(z) \hskip 1cm \mbox{and}  \hskip 1cm J(1/z)=1/J(z).
\label{z2b}
\ee
When applied to general degree three rational maps this can be used to restrict
to a one-parameter family of the form
\be
J(z)=\frac{cz^2-1}{z(z^2-c)}
\label{g3}
\ee
with $c$ complex. The easiest way to find the value of the constant $c$ for
tetrahedral symmetry is to examine the branch points of the rational map,
which must be invariant under the tetrahedral group. These are given
by the zeros of the numerator of the derivative
\be
\frac{dJ}{dz}=\frac{-c(z^4+(c-3/c)z^2+1)}{z^2(z^2-c)^2}.
\ee
By comparison of the numerator with the vertex and face Klein polynomials
of the tetrahedron ${\cal T}_{v,f}$, given by (\ref{kptvf}), 
it is clear that tetrahedral
symmetry results only if $c=\pm i\sqrt{3}.$ It can be checked that only
for these two values is the remaining $120^\circ$ rotation symmetry
of the tetrahedron attained as
\be
J\bigg(\frac{iz+1}{-iz+1}\bigg)=\frac{iJ(z)+1}{-iJ(z)+1}.
\label{s3}
\ee

In a similar manner the Jarvis maps of the other Platonic monopoles
can be constructed. These maps, together with many other symmetric maps,
can be found in ref.\cite{HMS}. In particular, a monopole corresponding to
the remaining Platonic solid, the icosahedron, is shown to exist for $k=11$,
as conjectured in ref.\cite{HS2}.

The simplicity of the construction of Jarvis maps for Platonic monopoles,
as compared to the computation of their Nahm data, is clearly evident.
However, if more detailed information regarding the monopole is sought,
such as the distribution of energy density or zeros of the Higgs field,
then the ADHMN construction is still the most efficient approach.
In principle it is possible to construct the monopole fields from the
Jarvis map\cite{Ja}, but this requires the solution of a nonlinear
partial differential equation that is just as difficult to solve
as the original Bogomolny equation (\ref{bog}). It is thus not as
effective as the ADHMN construction, which reduces the problem to
the solution of ordinary differential equations only. Nonetheless,
for cases in which the Nahm data is not known the construction
of monopoles from the Jarvis map is more appealing than a direct
solution of the Bogomolny equation, even when both need to be
implemented numerically, since it allows an elegant specification of exactly
which monopole is to be constructed.

\news
\vspace*{1pt}\textlineskip	
\section{Dynamics and Moduli Space Metrics}	
\vspace*{-0.5pt}
\noindent
So far we have discussed only static monopoles, which are solutions of the Bogomolny
equation (\ref{bog}). As we have seen this equation is integrable, 
which for the present case we shall take to mean that it has a twistor correspondence.
Unfortunately the full time-dependent field equations which follow from the
Lagrangian (\ref{lag}) are not integrable and so we do not expect to be able
to solve this equation explicitly, or apply a twistor transform, 
 to investigate monopole dynamics. However, progress can be made on the
study of slowly moving monopoles by applying Manton's moduli space approximation
(sometimes also called the geodesic approximation)\cite{Ma1}.

The moduli space approximation, which can also be applied to the dynamics of
other kinds of topological solitons\cite{Sa,Ru,Str,Wa5,Le}, was first proposed
for the study of monopoles\cite{Ma1}. Here one assumes that the $k$-monopole
configuration at any fixed time may be well-approximated by a static $k$-monopole
solution. The only time dependence allowed is therefore in the dynamics of the 
$4k$-dimensional $k$-monopole moduli space ${\cal M}_k.$ A Lagrangian on  ${\cal M}_k$
is inherited from the field theory Lagrangian (\ref{lag}), but since all
elements of  ${\cal M}_k$ have the same potential energy the kinetic part of the
action completely determines the dynamics. It defines a metric on  ${\cal M}_k$
and the dynamics is given by geodesic motion on  ${\cal M}_k$ with respect to this metric.

Intuitively one should think of the potential energy landscape in the 
 charge $k$ configuration space as having a flat valley given by ${\cal M}_k$,
and the low energy monopole dynamics takes place in, or at least close to,
this valley. A rigorous mathematical analysis of the validity of the
moduli space approximation has been performed by Stuart\cite{St}.

In the case of a single monopole the moduli space approximation is rather
trivial. ${\cal M}_1=\R^3\times S^1$ is flat, with constant motion in the
$\R^3$ giving the momentum of the monopole and constant angular speed in the $S^1$
determining the electric charge of the monopole, as mentioned in Section 1.

To study multi-monopole dynamics, as well as for other reasons which we
shall see later, it is therefore of interest to find the metric on ${\cal M}_k$.
By a formal application of the hyperk\"ahler quotient construction\cite{HKLR}
it can be shown\cite{AH} that ${\cal M}_k$ is a hyperk\"ahler manifold, which
means that there are three covariantly constant complex structures
satisfying the quaternionic algebra. The rotational symmetry of the system
means that there is also an $SO(3)$ action, which permutes the complex structures.
Physically, the motion of the centre of mass of the system and the total phase
decouples from the relative motion, which mathematically means that there
is an isometric splitting
\be \widetilde{{\cal M}}_k=\R^3\times\mbox{S}^1\times {\cal M}_k^0\ee
where $\widetilde{{\cal M}}_k$ is a $k$-fold covering of ${\cal M}_k$ and 
${\cal M}_k^0$ is the $4(k-1)$-dimensional moduli space of strongly centred $k$-monopoles.
The fact that a $k$-fold covering occurs is because the $k$ monopoles
are indistinguishable.

The non-trivial structure of the moduli space of 2-monopoles is therefore
contained in the totally geodesic 4-dimensional submanifold ${\cal M}_2^0.$
Using the fact that this is a hyperk\"ahler manifold with an $SO(3)$ symmetry,
Atiyah \& Hitchin\cite{AH} were able to reduce the computation of its metric to 
the solution of a single
ordinary differential equation, which can be done explicitly in terms of 
elliptic integrals.
They were thus able to explicitly determine the metric on ${\cal M}_2^0$, which
is now known as the Atiyah-Hitchin manifold. The easiest way to present the
Atiyah-Hitchin metric is in terms of the left-invariant 1-forms
\bea
\sigma_1&=&-\sin\psi d\theta+\cos\psi\sin\theta d\phi\\
\sigma_2&=&\cos\psi d\theta+\sin\psi\sin\theta d\phi\\
\sigma_3&=&d\psi +\cos\theta d\phi
\eea
where $\theta,\phi,\psi$ are the usual Euler angles.
It is given by
\be
ds^2=\frac{b^2}{K^2}dK^2+a^2\sigma_1^2+b^2\sigma_2^2+c^2\sigma_3^2
\label{ahmetric}
\ee
\bea
a^2&=&2K(K-E)(E-m'K)/E\\
b^2&=&2EK(K-E)/(E-m'K)\\
c^2&=&2EK(E-m'K)/(K-E)
\eea
where $K$ and $E$ denote the complete elliptic integrals of the first and second
kind with parameter $m$ and $m'=1-m$ is the complimentary parameter.

The coordinate $m$ determines the separation of the two monopoles in the same
way as appears in the 2-monopole spectral curve (\ref{sc2b}). That is, $m=0$
represents the axially symmetric 2-monopole (which is referred to as the
bolt in the metric context) and $m\rightarrow 1$ represents the
monopoles moving off to infinity.

There is an interesting totally geodesic 2-dimensional submanifold
 of the Atiyah-Hitchin manifold which can be obtained by
the imposition of a reflection symmetry. This 2-dimensional submanifold
is a surface of revolution, which metrically is a rounded cone\cite{AH}.
An interesting geodesic is one which passes directly over the cone ie. a
generator of the surface of revolution. In the moduli space approximation
this geodesic describes the head-on collision of two monopoles, which 
pass instantaneously through the axially symmetric 2-monopole and emerge
at $90^\circ$ to the initial direction of approach\cite{AH}. This 
famous right-angle scattering of monopoles has now been observed in
many other systems\cite{Sa,Ru,Str,Wa5,Le,LPZ,Su6} and appears to be a general
feature of multi-dimensional topological solitons.

Another interesting geodesic in the Atiyah-Hitchin manifold was discovered
by Bates \& Montgomery\cite{BM}. This is a closed geodesic and thus, within
the moduli space approximation, describes
a bound state of two orbiting monopoles.

The metric on ${\cal M}_k^0$ for $k>2$ is unknown and its difficulty
of computation is related to the earlier comment that there is an associated 
 algebraic curve of genus $(k-1)^2$, which is therefore no-longer simply elliptic.
However, some recent progress has been made regarding the computation of the
metric on certain totally geodesic submanifolds\cite{HS5,Bi2,BS} of ${\cal M}_k^0.$
In order to describe these results we now turn to a consideration of the
moduli space metric in terms of the Nahm transform.

The discussion of the ADHMN construction in Section 2 was presented in terms of
 Nahm data consisting of three Nahm matrices $(T_1,T_2,T_3)$, but in order to discuss the
metric we must, following Donaldson\cite{Do}, introduce a fourth
Nahm matrix $T_0$. Then we have that charge $k$ monopoles are 
equivalent to Nahm data $(T_0,T_1,T_2,T_3)$, which satisfy the full Nahm equations
\be
\frac{dT_i}{ds}+[T_0,T_i]=
\half\epsilon_{ijk}[T_j,T_k] \hskip 30pt i=1,2,3.
\label{fullnahm}
\ee
The Nahm data conditions remain the same as before, but are supplemented
by the requirement that $T_0$ is regular for $s\in[0,2].$

Let $H$ be the group of analytic $su(k)$-valued functions $h(s)$,
for $s\in[0,2]$, which are the identity at $s=0$ and $s=2$ and 
satisfy $h^t(2-s)=h^{-1}(s)$.
Then gauge transformations $h\in H$ act on Nahm data as
\be
T_0\mapsto hT_0h^{-1}-\frac{dh}{ds}h^{-1}, \hskip 1cm
T_i\mapsto hT_ih^{-1}  \hskip 15pt i=1,2,3. 
\ee

Note that the gauge $T_0=0$ may always be chosen, which is why 
this fourth Nahm matrix is usually not introduced. However, when
discussing the metric on Nahm data we need to consider the action of
the gauge group and so this extra Nahm matrix needs to be kept.

To find the metric on the space of Nahm data a basis for the tangent space
needs to be found, by solving the linearized form of Nahm's equation (\ref{fullnahm}).
Let $(V_0,V_1,V_2,V_3)$ be a tangent vector corresponding to the point 
with Nahm data $(T_0,T_1,T_2,T_3)$. It is a solution of the equations
\be
\frac{dV_i}{ds} +[V_0,T_i]+[T_0,V_i]=\epsilon_{ijk}[T_j,V_k]
\hskip 20pt i=1,2,3
\label{lne}
\ee
and
\be
\frac{dV_0}{ds}+\sum_{i=0}^3 [T_i,V_i]=0
\label{bg}
\ee
where $V_i,\ i=0,1,2,3$, is an analytic $su(k)$-valued function
of $s\in[0,2]$.
If $W_i$ is a second tangent vector then the metric component
corresponding to these two tangent vectors is defined as
\be
<V_i,W_i>=-\int_0^2\sum_{i=0}^3\mbox{tr}(V_iW_i)\ ds.
\label{ipnd}
\ee

Nakajima\cite{Nak1} has proved that the transformation between
the monopole moduli space metric and the metric on Nahm data
is an isometry. Therefore it is in principle possible to compute the metric
on  the monopole moduli space if the corresponding Nahm data is known.
For example, by using the known 2-monopole Nahm data, (\ref{nd2}) with (\ref{ndk2}),
the tangent vectors can be found explicitly in terms of elliptic
functions and the resulting integrals in (\ref{ipnd}) performed
in terms of complete elliptic integrals to recover the Atiyah-Hitchin
metric (\ref{ahmetric}).

Given the above discussion it is therefore of interest to look for families
of symmetric monopoles corresponding to totally geodesic submanifolds of
 ${\cal M}_k^0$, where the Nahm data can be found explicitly in terms of 
elliptic functions. We shall refer to such submanifolds as elliptic.
It is then possible that the metric on elliptic submanifolds
can be computed exactly, in terms of elliptic integrals. 

It can be shown\cite{HS5} that there is a 4-dimensional submanifold
of strongly centred 3-monopoles which are symmetric under the inversion
\hbox{${\bf x} \mapsto -{\bf x}$.} Since the fixed point set of a group
action is always totally geodesic, then this is a 4-dimensional
totally geodesic submanifold of ${\cal M}_3^0.$ The corresponding Nahm
data is a generalization of the 2-monopole Nahm data (\ref{nd2}) where
the same functions (\ref{ndk2}) occur but the matrices which form
the spin $\half$ representation of $su(2)$ are replaced by
those of the spin 1 representation. The upshot of this is that the
only modification which arises in the computation of the metric from
the Nahm data is the overall  multiplication by a constant. Thus,
after checking that rotations act in the same way as before, this
proves\cite{HS5} that this submanifold is a totally geodesic Atiyah-Hitchin
submanifold of ${\cal M}_3^0.$ 

Physically, the three monopoles are collinear, with the third monopole
fixed at the origin and the other two behaving in an Atiyah-Hitchin
manner, as if they do not see this third one. Simultaneously with this
 discovery Bielawski\cite{Bi2} found the same result using the 
hyperk\"ahler quotient construction and also showed that it generalizes
to a totally geodesic Atiyah-Hitchin submanifold of ${\cal M}_k^0$,
for all $k>2$, corresponding to a string of $k$ equally spaced
collinear monopoles. The Nahm data in each case is again obtained
by replacing the spin $\half$ representation of $su(2)$ by
the spin $(k-1)/2$ representation, but the submanifold can not
be obtained as the fixed point set of a group action. However, the
fact that it is totally geodesic can be deduced from the
knowledge that it is a hyperk\"ahler submanifold of a hyperk\"ahler manifold.
Note that once the manifold is shown to have an Atiyah-Hitchin
submanifold then it is guaranteed to possess a closed geodesic
describing a bound state of $k$ monopoles, since it is merely
the inherited Bates \& Montgomery geodesic in ${\cal M}_2^0.$

A geodesic in ${\cal M}_4^0$ can be obtained by the imposition of
tetrahedral symmetry on 4-monopoles\cite{HS1}. This 1-dimensional
submanifold is elliptic, allowing the computation of the associated
Nahm data, which can be used to study the scattering of 4-monopoles\cite{HS1}
 and also the calculation of the 1-dimensional metric\cite{BS,Su1}.
This family obviously includes the cubic 4-monopole and thus via
the moduli space approximation it describes the scattering of four monopoles
which pass through the cubic configuration. In Fig.2 we display surfaces
of constant energy density at increasing times throughout this process.
It can be seen that four monopoles approach from infinity on the
vertices of a contracting regular tetrahedron, coalesce to form a
configuration with instantaneous cubic symmetry and emerge
on the vertices of an expanding tetrahedron dual to the incoming one.
\begin{figure}[htbp]
\ \vskip -3cm
\epsfxsize=12cm \epsffile{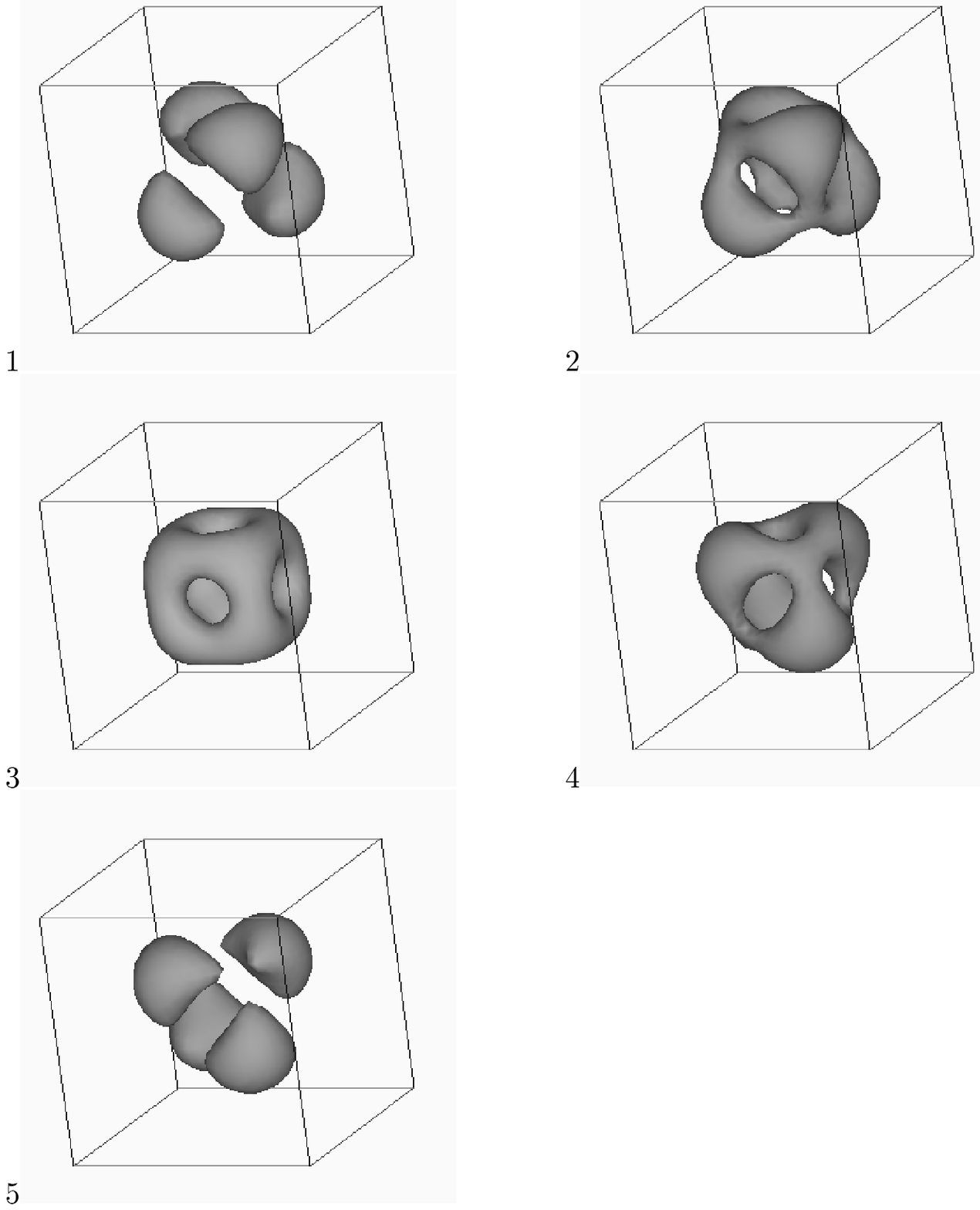}
\ \vskip -6cm
\fcaption{Energy density surfaces for tetrahedral 4-monopole scattering.}
\end{figure}
Note that in the above example it is not necessary to know the metric
in order to determine the trajectories of the monopoles. This is because
an application of a symmetry resulted in a totally geodesic 1-dimensional
submanifold, which by definition is a geodesic. Thus the metric only
determines the rate at which motion takes place along the geodesic
and is not needed to determine the form of the motion.

This observation is useful and suggests the scheme of searching
for appropriate symmetry groups $G$, for various charges $k$, so that
the moduli space of $G$-symmetric $k$-monopoles is 1-dimensional, thus
furnishing a geodesic.

Searching for symmetric monopoles using the ADHMN construction is not
a simple task, since in each case it relies upon a careful study
of solutions to Nahm's equation. Furthermore, unless the submanifold
in question is elliptic, it is unlikely that this will be a tractable problem.
However, using rational maps it is a much easier problem to analyse.

Recall that Donaldson rational maps require a choice of direction in $\R^3$,
so only symmetries which preserve this direction can be studied using these
maps. Nonetheless, it turns out that this still provides a substantial
collection of interesting geodesics. 

The obvious symmetry to impose which fixes a direction is cyclic symmetry
and the corresponding rational maps were investigated by 
Hitchin, Manton \& Murray\cite{HMM}.
Requiring invariance of
a $k$-monopole under cyclic
$C_k$ symmetry and an additional reflection symmetry, leads to
a number of geodesics, $\Sigma_k^l$, 
 in the $k$-monopole moduli space ${\cal M}_k$.
Essentially there are $(2k+3+(-1)^k)/4$ different types of
these geodesics, corresponding to $l=0,1,..k/2$ if $k$ is even
and $l=0,1,..(k-1)/2$ if $k$ is odd. Physically, for $l\ne 0$,
the associated
monopole scatterings are distinguished by having the out state
(or in state by time reversal) consisting of two clusters of
monopoles with charges $k-l$ and $l$. This explains why we do not
allow $l>k/2$, since this is basically the same scattering event as one
of the geodesics with $l<k/2$. 
If $l=0$ then the monopoles remain in a plane and scatter
instantaneously through the axisymmetric $k$-monopole and emerge
with a $\pi/k$ rotation. In this case if $k=2$ then this is just the Atiyah-Hitchin
right-angle scattering that we have already discussed. 
In fact the case $k=2$ is special, in that the geodesics $\Sigma_2^0$ and
$\Sigma_2^1$ are isomorphic, so that there is only this one type
of scattering. 
For all $k$ with $l=0$ this kind of $\pi/k$ scattering is
essentially a two-dimensional process and has been extensively
studied in planar systems\cite{KPZ}.

For $l\ne 0$ we see that the scatterings are more exotic, since the 
clustering of monopoles changes during the scattering process.
It can be shown\cite{Su2} that for these cases Nahm's equation reduces
to the $A_{k-1}^{(1)}$ Toda chain and the quotient spectral curve
has genus $k-1.$ Again this curve is not elliptic (except for the
case $k=2$ when the Toda chain is equivalent to the static sine-Gordon equation
with the appropriate solution being the kink)
but perhaps relating the ADHMN construction to other well-studied
integrable systems may prove fruitful. In fact other aspects of the
ADHMN construction can also be connected with more traditional
integrable systems, such as Lam\a'e\ equations\cite{Wa3,Su3},
and it would be worth while exploring these relationships further.

Even though the Nahm data for these cyclic monopoles has not
been found explicitly, it is possible to find an approximation
to it and hence display the scattering events\cite{Su4}. Fig.3 shows
one such example for the geodesic $\Sigma_3^1.$ It shows three
individual monopoles which lie on the vertices of a contracting equilateral
triangle in the plane, which merge to instantaneously form the tetrahedral
3-monopole and finally split to form a charge two torus and a single monopole
moving apart along an axis orthogonal to the plane of the incoming monopoles.
\begin{figure}[htbp]
\epsfxsize=12cm \epsffile{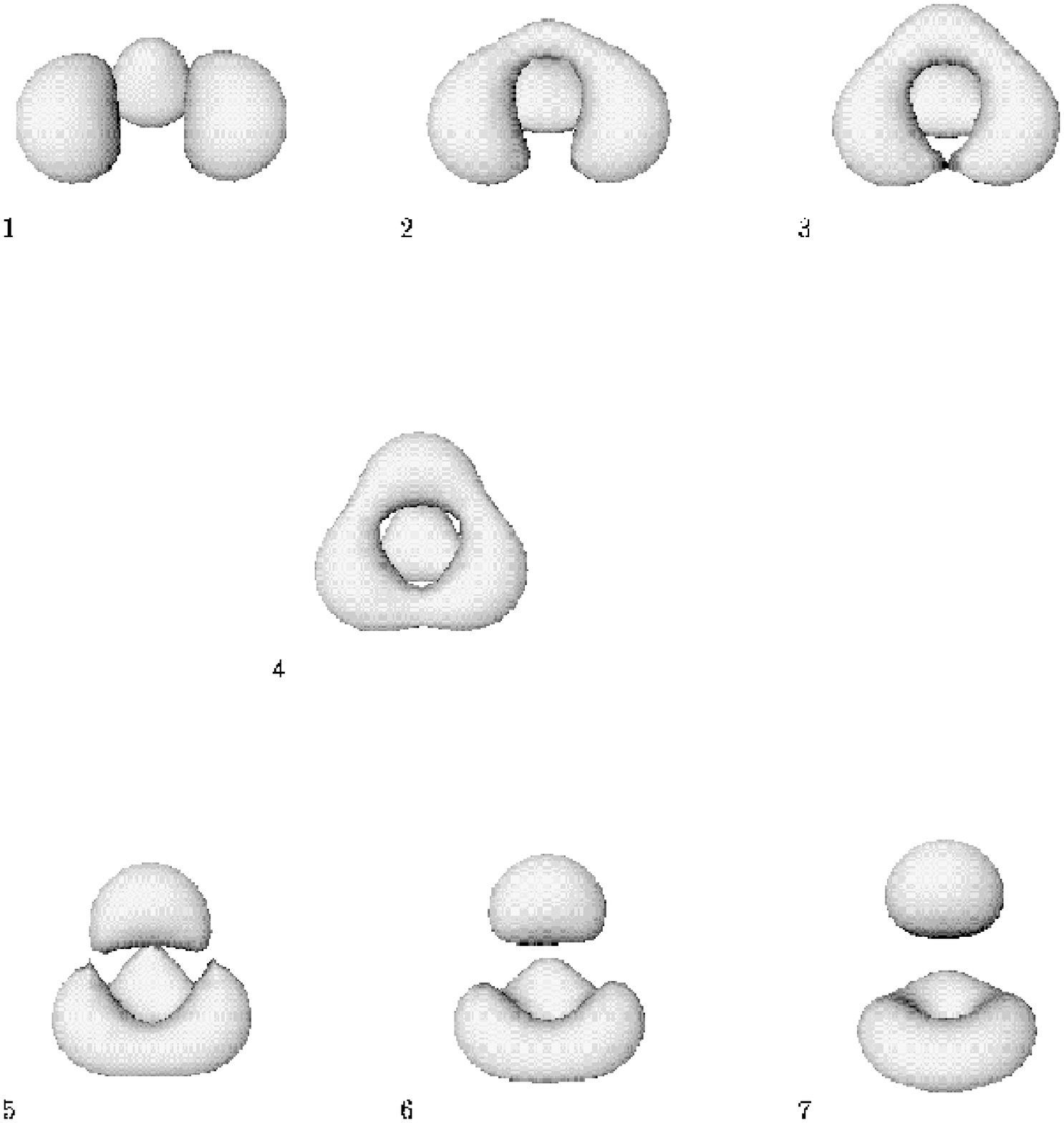}
\fcaption{Energy density surfaces for cyclic 3-monopole scattering.}
\end{figure}

Another class of symmetries which it is worth investigating 
 using Donaldson maps are the twisted cyclic symmetries obtained by the 
composition of a rotation with a reflection in a plane orthogonal to
 the axis of rotation. Geodesics can be obtained in this way\cite{HS3} which 
describe monopoles scattering through all the Platonic configurations discussed
earlier. In the simplest case of 3-monopoles with a twisted $90^\circ$ rotation
symmetry the corresponding geodesic is elliptic and the Nahm data has
been computed exactly. Explicitly the one-parameter family of spectral curves
is given by 
\be
\eta^3-6(a^2+4\epsilon)^{1/3}\kappa^2\eta\zeta^2
+2i\kappa^3a(\zeta^5-\zeta)=0,
\ee
 ($\epsilon=\pm1$), where $\kappa$ is the real half-period of the elliptic
curve
\be
y^2=4x^3-3(a^2+4\epsilon)^{2/3}x+4\epsilon.
\label{ellipc}
\ee
This includes the tetrahedral 3-monopole and its dual ($a=\pm 2,\epsilon=-1$),
the axisymmetric 3-monopole ($a=0,\epsilon=-1$) with symmetry axis $x_3$, 
and asymptotically ($a\rightarrow 0, \epsilon=1$)
describes three collinear monopoles which lie along the $x_3$-axis.

As mentioned earlier, some of the above included configurations have
spurious anti-zeros of the Higgs field, while some do not, such as
the well-separated limit. Thus there must be special \lq splitting points\rq\
at which anti-zeros appear, or disappear. In this example it seems that
these \lq splitting points\rq\ occur at $a=\pm\sqrt{8},\epsilon=-1$
and $a=0,\epsilon=-1$, which are all the points where
the discriminant of the elliptic curve (\ref{ellipc}) vanishes, so that
the curve is rational. The motion of the zeros and anti-zeros throughout 
this scattering is discussed in detail in ref.\cite{HS3}.

The recently introduced Jarvis maps, which we reviewed in Section 2,
 are much better for identifying geodesics obtainable by symmetry considerations.
Some examples are given in ref.\cite{HMS}, such as an interesting
scattering process involving seven monopoles which pass through two dodecahedra
and a cube.

Although the full metric on ${\cal M}_k^0$ is unknown for $k>2$ it is
known asymptotically on regions which correspond to all the $k$ monopoles
being well-separated\cite{GM2}.

First consider the Atiyah-Hitchin metric (\ref{ahmetric}) for large
monopole separations. The asymptotic form can be found by using the
standard expansions for elliptic integrals as $m\rightarrow 1.$
Writing $\rho=2K$, which asymptotically is the separation of the two monopoles,
it is given by
\be
ds^2=\bigg(1-\frac{2}{\rho}\bigg)(d\rho^2+\rho^2d\theta^2+\rho^2\sin^2\theta d\phi^2)
+4\bigg(1-\frac{2}{\rho}\bigg)^{-1}(d\psi+\cos\theta d\phi)^2.
\label{asy2}
\ee
This is the Taub-NUT metric with a negative mass parameter and is also
a  hyperk\"ahler metric. It has a singularity
at $\rho=2$ but this is not a value for 
which the approximation is valid, since it assumes that $\rho\gg 1.$
Note that (\ref{asy2}) also has a $U(1)$ symmetry which the Atiyah-Hitchin
metric lacks and corresponds physically to the conservation of 
 the relative electric charge. Asymptotically the metric is correct
up to terms which are exponentially suppressed.

Manton\cite{Ma6} pointed out that this asymptotic metric could be derived
from a point particle approximation, by treating each monopole as a source
of electric, magnetic and scalar charge. A similar calculation for the general
charge $k$ case was performed by Gibbons \& Manton\cite{GM2} 
and results in the following.
For $k$ monopoles located at $\{\rhobf_{i}\}$ 
with phases $\{\theta_i\}$ the asymptotic metric is
\be ds^2=g_{ij}d\rhobf_i\cdot d\rhobf_j+g_{ij}^{-1}(d\theta_i+{\bf
  W}_{ik}\cdot d\rhobf_k)(d\theta_j+{\bf
  W}_{jl}\cdot d\rhobf_l)\ee
where $\cdot$ denotes the usual scalar product on $\R^3$ vectors,
repeated indices are summed over and
\bea g_{jj}&=&1-\sum_{i\not=j}\frac{1}{\rho_{ij}}\qquad\mbox{(no sum over
  }j\mbox{)}\\
     g_{ij}&=&\frac{1}{\rho_{ij}}\,\;\qquad\mbox{(}i\not=j\mbox{)}\nonumber\\
     {\bf W}_{jj}&=&-\sum_{i\not=j}{\bf w}_{ij}\,\,\;\qquad\mbox{(no sum over
  }j\mbox{)}\nonumber\\
     {\bf W}_{ij}&=&{\bf w}_{ij}\,\,\qquad\mbox{(}i\not=j\mbox{)},\nonumber\eea
$\rhobf_{ij}=\rhobf_i-\rhobf_j$ and $\rho_{ij}=|\rhobf_{ij}|$. The
approximation is valid for
$\rho_{ij}\gg1$. The ${\bf w}_{ij}$ are Dirac potentials and are defined
by 
\be \mbox{curl}\ {\bf w}_{ij}=\mbox{grad} \frac{1}{{ \rho}_{ij}}\ee 
where the curl and grad operators are taken with respect to the $i$th
position coordinate $\rhobf_i$.

It can be checked, for example, that the metric on the moduli space
of tetrahedrally symmetric charge four monopoles, which is known
exactly in terms of complete elliptic integrals\cite{BS}, agrees
with this formula asymptotically.

\news
\vspace*{1pt}\textlineskip	
\section{Higher Rank Gauge Groups}	
\vspace*{-0.5pt}
\noindent
So far we have dealt only with the case of $SU(2)$ monopoles.
The kind of analysis we have reviewed can, of course, be extended
to more general gauge groups, where things usually become more complicated.
In this Section we sketch how the ideas and results are modified for
$SU(N)$ gauge groups and
discuss some special situations in which the problem simplifies.

Recall from Section 1 that in a
gauge theory where the non-abelian gauge group $G$ is spontaneously broken
by the Higgs field to a residual symmetry group $H$ then the monopoles
have a topological classification determined by the elements of
$\pi_2(G/H).$

For $G=SU(N)$ then the boundary conditions at spatial infinity are that
 $\Phi$ takes values in the
gauge orbit of the matrix
\be M=i\,\mbox{diag}\,(\mu_1,\mu_2,\ldots,\mu_N).\ee
By convention it is assumed that $\mu_1\le\mu_2\le\ldots\le\mu_N$ and since
$\Phi$ is traceless then $\mu_1+\mu_2+\ldots+\mu_N=0$. This $M$ is the
vacuum expectation value for $\Phi$ and the residual symmetry group $H$ is the
symmetry group of $M$ under gauge transformations.
Thus, for example, if all the $\mu_p$ are
distinct then the residual symmetry group is the maximal torus
$U(1)^{N-1}$ and this is known as maximal symmetry breaking.
In this case
\be
\pi_2\left(\frac{SU(N)}{U(1)^{N-1}}\right)=\pi_1(U(1)^{N-1})=\Z^{N-1}
\ee
so the monopoles are associated with $N-1$ integers.

In contrast, the minimal symmetry breaking case is that 
 in which all but one of the $\mu_p$ are identical, so the
residual symmetry group is $U(N-1)$.
Since
\be
\pi_2\left(\frac{SU(N)}{U(N-1)}\right)=\Z
\ee 
there is only one topological integer characterization of a monopole. 
Nonetheless, a given solution has $N-1$ integers associated
with it, which arise in the following way.

 A careful analysis of the
boundary conditions\cite{GNO,We2} indicates that there is a choice of gauge such
that the Higgs field for large $r$, in a given direction, is given by
\be
\Phi(r)=i\,\mbox{diag}\,(\mu_1,\mu_2,\ldots,\mu_N)-\frac{i}{r}\,
\mbox{diag}\,(k_1,k_2,\ldots,k_N)+O(r^{-2}).
\ee
In the maximal symmetry breaking case the
topological charges are given by 
\be m_p=\sum_{q=1}^pk_q.\ee
In the case of minimal symmetry breaking only the first of these
numbers, $m_1$, is a topological charge. Nonetheless, the remaining
$m_p$ constitute an integer characterization of a solution, which
is gauge invariant up to reordering of the
integers $k_p$. The $m_p$ are known as magnetic weights, with the matrix
 $\mbox{diag}\,(k_1,k_2,\ldots,k_N)$ often called the
charge matrix and
$\mbox{diag}\,(\mu_1,\mu_2,\ldots,\mu_N)$ the mass matrix.

There are some obvious ways of embedding $su(2)$ in $su(N)$, for example,
\be \left(\begin{array}{cc}\alpha&\beta\\-\bar{\beta}&-\alpha\end{array}\right)\hookrightarrow\left(\begin{array}{ccccc}\ddots&&&&\\ &\alpha&\ldots&\beta& \\
&\vdots&\ddots&\vdots&\\ &-\bar{\beta}&\ldots&-\alpha& \\&&&&\ddots\end{array}\right).\label{emb}\ee
Important $SU(N)$ monopoles can be produced by
embedding the $SU(2)$ charge one monopole fields, which are known $su(2)$-valued fields, in
$su(N)$. Some care must be taken in producing these embedded monopoles to ensure
that the asymptotic behaviour is correct. The $SU(2)$ monopole may
need to be scaled and it may be necessary to add a
constant diagonal field beyond the plain embedding described by
(\ref{emb}); details may be found in refs.\cite{We2,Wa6}. Obviously there is an
embedding of the form (\ref{emb}) for each choice of two columns in
the target matrix. The embedded 1-monopoles have a single $k_p=1$ and
another $k_{p'}=-1$, the rest are zero. The choice of columns for
the embedding dictates the values for $p,p'$, so there are $N-1$ different
types of fundamental monopole with $m_p=1$ and the rest zero, corresponding to
the choice $p'=p+1.$

Recall that in the case of minimal symmetry breaking the choice of order of
the $k_p$ is a gauge choice.
In fact, in the case of minimal symmetry breaking, the embedded
1-monopole is unique up to position and gauge
transformation. Solutions with $k_1=k$ have $k$ times the energy of this
basic solution and so it is reasonable to call these 
$k$-monopoles. There are of course different types of such $k$-monopoles
corresponding to different magnetic weights.

For other intermediate cases of symmetry breaking 
 the residual symmetry group is $H=U(1)^r\times K$,
where $K$ is a rank $N-r-1$ semi-simple Lie group, the exact form of which
depends on how the entries in the mass matrix coincide with each other.
Such monopoles have $r$ topological charges.

The twistor methods of Section 2 can be formulated for the case
of general gauge groups. Ward\cite{Wa6} has constructed some explicit
$SU(3)$ monopoles via the splitting of appropriate patching matrices over
\mt. The spectral curve approach for maximal symmetry breaking has been formulated by 
Hurtubise \& Murray\cite{HM} and consists of a specification of 
$\mbox{rank}(G)$ algebraic curves in \mt, satisfying reality and non-singularity
conditions. For
higher rank gauge groups the Donaldson rational map correspondence
has been extended by Murray\cite{Mu1}
to maps into flag manifolds and
a similar extension exists for the new rational maps of Jarvis\cite{Ja}.

 The ADHMN construction for general $G$ is outlined in the original work
of Nahm\cite{Na} and is discussed further in ref.\cite{HM}. Briefly,
for $G=SU(N)$ the Nahm data are triplets of anti-hermitian matrix functions
$(T_1,T_2,T_3)$ of $s$ over the intervals $(\mu_p,\mu_{p+1})$. The size of
the matrices depends on the corresponding values of $m_p$; the
matrices $(T_1,T_2,T_3)$ are $m_p\times m_p$ matrices in the interval
$(\mu_{p},\mu_{p+1})$. They are required to be non-singular inside each
interval and to satisfy  Nahm's equation (\ref{nahms})
but there are complicated boundary conditions at the ends of each of the intervals.
These boundary conditions are designed so that
the linear equation (\ref{lin}) has the correct number of solutions required
to yield the right type of monopole fields.

The simplest case is maximal symmetry breaking in an $SU(3)$ theory.
There are then two types of monopole, so the charge is a 2-component vector
$(m_1,m_2).$ The simplest multi-monopole is therefore of charge $(1,1)$,
and its Nahm data was studied by Connell\cite{Co}. Since there is only one
of each type of monopole then the Nahm data is 1-dimensional over each
of the two intervals, so Nahm's equations are trivially satisfied by
taking the Nahm data to be constants over each of the two intervals.
These two triplets of constants determine the positions of the two monopoles
and the matching condition at the common boundary of the two intervals
determines the relative phase.

The moduli space of these monopoles is 8-dimensional but, as in the $SU(2)$ case,
there is an isometric splitting to factor out the position of the centre of
mass and the overall phase. The relative moduli space, ${\cal M}_{(1,1)}^0$,
is thus 4-dimensional. By computing the metric on Nahm data and using
a uniqueness argument, Connell\cite{Co} was able to show that the metric on
${\cal M}_{(1,1)}^0$ is the Taub-NUT metric with a positive mass parameter.
This result was rediscovered some years later\cite{LWY1,GL}. Recall that
the asymptotic Atiyah-Hitchin metric is also a Taub-NUT metric, but with a negative
mass parameter, so the asymptotic metric has a singularity 
outside its region of validity. This difference in sign results from the fact
that in the $SU(3)$ case the two monopoles are electrically charged with respect
to different $U(1)$ factors in the residual symmetry group. There is thus
a conservation of the individual electric charge of each monopole, providing
a $U(1)$ symmetry in the metric which is absent in the Atiyah-Hitchin metric,
since charge exchange occurs between $SU(2)$ monopoles. This results is a simplified
dynamics of charge $(1,1)$ monopoles, which bounce back off each other in a head-on
collision in comparison with the right-angle scattering of two $SU(2)$ monopoles.

Similar simplifications can be expected in all cases where there is at most
a single monopole of each type. Thus the $4(N-2)$-dimensional relative moduli space
$\widehat{\cal M}_N^0$ of charge $(1,1,...,1)$ monopoles in an $SU(N)$
theory should be tractable. Indeed, Lee, Weinberg \& Yi\cite{LWY2} have computed
the asymptotic metric, which is a generalization of the Taub-NUT case,
and conjectured that it is the exact metric. This is supported by a computation
of the metric on the space of Nahm data by Murray\cite{Mu2}, which obtains the
same result. Note that this last calculation is not quite a proof, since
although it is believed that the transformation between the monopole moduli
space metric and the metric on Nahm data is an isometry for all gauge groups
and symmetry breaking, it has only been proved for $SU(2)$ monopoles\cite{Nak1}
and special cases for minimally broken $SU(N)$\cite{NT}. More recently,
these and other monopole metrics have also been obtained by 
Gibbons \& Rychenkova\cite{GR} using the hyperk\"ahler quotient construction.

There is a method which can be used to give a local construction of
  hyperk\"ahler metrics known as the generalized Legendre transform\cite{HKLR}.
This can be used, for example, to give yet another derivation\cite{IR} of 
the Atiyah-Hitchin manifold. Using this method Chalmers\cite{Ch1} was
able to rederive the Lee-Weinberg-Yi metric and prove that it is the correct
metric throughout the moduli space.

In order to examine if there are any other special choices of gauge group,
symmetry breaking and monopole charges for which there may be a simplification
we need to review a few more details of the Nahm data boundary conditions.

For ease of notation we shall only describe the case where $m_{p-1}> m_p$,
since this will be the one of interest in what follows.
Define the function
\be k(s)=\sum_{p=1}^Nk_p\,\theta(s-\mu_p)\ee
where $\theta(s)$ is the usual step function. In the interval
$(\mu_p,\mu_{p+1})$ then $k(s)=m_p$, so it is a rectilinear
skyline whose shape depends on the charge matrix of the corresponding monopole. 

\newpage
If $k(s)$ near $\mu_p$ looks like\\

{
\begin{center}
\setlength{\unitlength}{0.012500in}%
\begingroup\makeatletter\ifx\SetFigFont\undefined
\def\x#1#2#3#4#5#6#7\relax{\def\x{#1#2#3#4#5#6}}%
\expandafter\x\fmtname xxxxxx\relax \def\y{splain}%
\ifx\x\y   
\gdef\SetFigFont#1#2#3{%
  \ifnum #1<17\tiny\else \ifnum #1<20\small\else
  \ifnum #1<24\normalsize\else \ifnum #1<29\large\else
  \ifnum #1<34\Large\else \ifnum #1<41\LARGE\else
     \huge\fi\fi\fi\fi\fi\fi
  \csname #3\endcsname}%
\else
\gdef\SetFigFont#1#2#3{\begingroup
  \count@#1\relax \ifnum 25<\count@\count@25\fi
  \def\x{\endgroup\@setsize\SetFigFont{#2pt}}%
  \expandafter\x
    \csname \romannumeral\the\count@ pt\expandafter\endcsname
    \csname @\romannumeral\the\count@ pt\endcsname
  \csname #3\endcsname}%
\fi
\fi\endgroup
\begin{picture}(360,121)(100,524)
\thinlines
\put(300,540){\line( 0,-1){  6}}
\put(330,600){\vector( 0, 1){  0}}
\put(330,600){\vector( 0,-1){ 55}}
\put(250,640){\vector( 0, 1){  0}}
\put(250,640){\vector( 0,-1){ 94}}
\multiput(403,605)(8.22222,0.00000){5}{\line( 1, 0){  4.111}}
\put(180,645){\line( 1, 0){120}}
\put(300,645){\line( 0,-1){ 40}}
\put(300,605){\line( 1, 0){100}}
\put(400,605){\line(-1, 0){  1}}
\put(100,540){\line( 1, 0){380}}
\multiput(179,645)(-8.66667,0.00000){5}{\line(-1, 0){  4.333}}
\put(314,624){\makebox(0,0)[lb]{\smash{\SetFigFont{12}{14.4}{rm}$-k_p$}}}
\put(310,645){\vector( 0, 1){  0}}
\put(310,645){\vector( 0,-1){ 36}}
\put(286,524){\makebox(0,0)[lb]{\smash{\SetFigFont{12}{14.4}{rm}$s=\mu_p$}}}
\put(254,590){\makebox(0,0)[lb]{\smash{\SetFigFont{12}{14.4}{rm}$m_{p-1}$}}}
\put(335,570){\makebox(0,0)[lb]{\smash{\SetFigFont{12}{14.4}{rm}$m_p$}}}
\put(445,602){\makebox(0,0)[lb]{\smash{\SetFigFont{12}{14.4}{rm}$k(s)$}}}
\end{picture}
\end{center}
}
then as $s$ approaches $\mu_p$ from below it is required that
\be \begin{array}{c}\\T_i(s)=\end{array}\begin{array}{c}|k_p|\quad\,\,\,\,\,\qquad\qquad\; m_p\,\qquad
\\ \left(\begin{tabular}{c|c}&\\[-10pt]$\frac{1}{z}R_i+O(1)$ &
$O(z^{(|k_p|-1)/2})$ \\[10pt] \hline \\[-10pt]$O(z^{(|k_p|-1)/2})$   & $T_i^{\prime}+O(z)$\\[-15pt]& \end{tabular}\right)\begin{array}{c}\\[-9pt]|k_p|\\[15pt]m_p\end{array}\end{array}\label{frombelow}\ee
where $z=s-\mu_p$ and where
\be
T_i(s)=T_i^{\prime}+O(z)\ee
as $s$ approaches $\mu_p$ from above.
It follows from  Nahm's equation (\ref{nahms}) that the
 $|k_p|\times |k_p|$ residue matrices $(R_1,R_2,R_3)$ in
(\ref{frombelow}) form a representation of $su(2)$. The boundary
conditions require that this representation is the unique irreducible
$|k_p|$-dimensional representation of $su(2)$. 

In summary, at the boundary between
two intervals, if the Nahm matrices are $m_{p-1}\times m_{p-1}$ on the
left and $m_{p}\times m_{p}$ on the right  an
$m_p\times m_p$ block continues through the boundary and there is an
$(m_{p-1}-m_p)\times(m_{p-1}-m_p)$ block simple pole whose residues
form an irreducible representation of $su(2)$.

These conditions now suggest a simplifying case, since if $k_p=-1$ for
all $p>1$ then $k(s)$ is a staircase with each step down of unit height.
We shall refer to this situation as the countdown case since
the magnetic weights are given by $(N-1,N-2,...,2,1).$
Thus, since all the 1-dimensional representations of $su(2)$ are trivial,
the Nahm data has only one pole, which is at $s=\mu_1.$
Taking the limiting case of minimal symmetry breaking, by setting
$\mu_1=-(N-1)$ and $\mu_2=...=\mu_N=1$, we find that the Nahm data
is defined on a single interval $[-N+1,1]$ with the only pole occurring
at the left-hand end of the interval. This is very similar to the Nahm data
for $SU(2)$ monopoles, except that the pole at the right-hand end of the interval 
is lost. This allows a construction of Nahm data for charge $N-1$ monopoles
in a minimally broken $SU(N)$ theory in terms of rescaled Nahm data
for $SU(2)$ monopoles, where the rescaling moves the second pole in the Nahm data
outside the interval. With this in mind it is convenient to shift $s$ so that
the Nahm data is defined over the interval $[0,N].$

As an illustration of the simplification that occurs in the countdown case
we present the Nahm data for an $SU(N)$ charge $N-1$ spherically symmetric 
monopole.
It is given by $T_i=-\rho_i/2s$ where $\rho_1,\rho_2,\rho_3$ form the standard
spin $(N-2)/2$ representation of $su(2).$ The associated spectral curves are
simply $\eta^{N-1}=0.$ Spherically symmetric monopoles were first studied
by Bais \& Wilkinson\cite{BW}, Leznov \&  Saveliev\cite{LS} 
and Ganoulis,  Goddard \&  Olive\cite{GGO}, all using a radial ansatz in the
 Bogomolny equation.

The simplest countdown example to consider further is the case of 
charge two $SU(3)$ monopoles with
 minimal symmetry breaking. For $k=2$
there are two distinct types of monopoles
corresponding to magnetic weights $(2,0)$ and $(2,1).$ (The cases $(2,2)$ and $(2,0)$ are
equivalent by a reordering of $k_2$ and $k_3$). For weights $(2,0)$ the
monopoles are all embeddings of $su(2)$ 2-monopoles and this case is not
interesting as an example of $su(3)$ 2-monopoles. For weights $(2,1)$ this
is a countdown case and was
first studied by Dancer\cite{Da1,Da2}. Given the comments above it is
fairly clear that  the appropriate Nahm data has the same form
as the $SU(2)$ 2-monopole Nahm data (\ref{nd2}) where the functions
$f_1,f_2,f_3$ are almost the same as in the $SU(2)$ case (\ref{ndk2})
except that the complete elliptic integral $K$, whose value was required
to place the second pole at $s=2$, is now replaced by a parameter $D$,
whose range is such that no second pole occurs in the interval ie.
$D<2K/3.$ Explicitly the Nahm data is
\be
 T_1=-i\frac{D\mbox{dn}(Ds)}{2\mbox{sn}(Ds)}\sigma_1,\ \
 T_2=-i\frac{D}{2\mbox{sn}(Ds)}\sigma_2, \ \
T_3=i\frac{D\mbox{cn}(Ds)}{2\mbox{sn}(Ds)}\sigma_3
 \label{dancer}
\ee

The moduli space of such monopoles
is 12-dimensional, so after centering we are left with an 8-dimensional
relative moduli space $M_8^0.$ There is an isometric 
$SO(3)\times SU(2)/\Z_2$
action on $M_8^0$. The $SU(2)/\Z_2$ action is a gauge transformation
on the four Nahm matrices, which is the identity at $s=0$, while the
$SO(3)$ action both rotates the three Nahm matrices as a vector
and gauge transforms all four Nahm matrices\cite{Da1}. 
Taking the quotient of $M_8^0$ by the $SU(2)/\Z_2$ action gives a 5-dimensional
manifold $M_5^0$ which has an $SO(3)$ action, since the $SU(2)/\Z_2$ and $SO(3)$
actions on $M_8^0$ commute. The Nahm data for $M_5^0$ is precisely the
orbit under $SO(3)$ of the 2-parameter family of Nahm data (\ref{dancer}).
Using this Nahm data, Dancer\cite{Da1} computed an explicit expression
for the metric on  $M_5^0$ and an implicit form for the metric on the
whole of $M_8^0.$ A more explicit form for the metric on $M_8^0$, in
terms of invariant 1-forms corresponding to the two group actions,
together with a study of the corresponding asymptotic monopole fields
has been given by Irwin\cite{Ir}.

A totally geodesic 2-dimensional submanifold, $Y$, of $M_8^0$ is obtained
by imposition of a $\Z_2\times\Z_2$ symmetry, representing monopoles
which are symmetric under reflection in all three Cartesian axes.
In fact $Y$ consists of six copies of the space $M_5^0/SO(3).$
This submanifold was introduced by Dancer \& Leese and the geodesics
and corresponding monopole dynamics investigated\cite{DL1,DL2}.
There are two interesting new phenomena which occur. The first is that
there can be double scatterings, where the two monopoles scatter at
right-angles in two orthogonal planes. The second is that there are
unusual geodesics which describe monopole dynamics where 
two monopoles approach from infinity
but stick together, with the motion taking the configuration
asymptotically towards an embedded $SU(2)$ field, which is on the boundary
of the $SU(3)$ monopole moduli space and metrically at infinity.
This kind of behaviour is still not completely understood but the
interpretation is that there is a non-abelian cloud\cite{LWY3,Ir},
whose radius is related to the parameter $D$ in the Nahm data (\ref{dancer}).
It is the motion of this cloud which carries off the kinetic energy
when the monopoles stick. Lee, Weinberg \& Yi\cite{LWY3} interpret this
cloud as the limit of a charge $(2,1)$ monopole in a maximally broken
theory, in which the mass of the $(\ ,1)$ monopole is taken to zero, thereby
losing its identity and becoming the cloud.

For the case of charge $(2,1)$ monopoles in the maximally broken
$SU(3)$ theory, Chalmers has conjectured an implicit form
for the metric\cite{Ch1}. This uses the generalized Legendre transform
technique and modifies the same construction of the Atiyah-Hitchin metric\cite{IR}.

Nahm data for other $SU(N)$ countdown examples can be obtained by a modification
of $SU(2)$ Nahm data. For example, Platonic $SU(N)$ monopoles can be studied
from the $SU(2)$ Nahm data discussed in Section 3. Again exotic phenomena are
found such as double scatterings and pathological geodesics where the
monopoles never separate\cite{HS6}.

\news
\vspace*{1pt}\textlineskip	
\section{S-Duality and Seiberg-Witten Theory}	
\vspace*{-0.5pt}
\noindent
There has been a recent revival of  interest in BPS monopoles due
to their central role in S-duality and Seiberg-Witten theory.
It is beyond the scope of this review to discuss these topics in any detail
but a few remarks regarding the application of results in monopole theory
 will be made.

Montonen-Olive duality\cite{MO} is a conjectured weak to strong coupling $\Z_2$ symmetry 
 which exchanges electric and magnetic charge and thus interchanges the
fundamental particles and monopoles. It was soon realised\cite{Os} that the best
chance of this duality existing occurs in an ${\cal N}=4$ supersymmetric Yang-Mills theory.
Much later is was observed, following work on string theory and lattice models, that
this $\Z_2$ symmetry should be extended to a conjectured $SL(2,\Z)$ symmetry, which
is now known as S-duality. The real explosion in this subject took place with the
work of Sen\cite{Se} who showed that a consequence of S-duality is the existence
of certain monopole-fermion bound states, which are described by self-dual normalizable
harmonic forms on the classical centred monopole moduli space ${\cal M}_k^0.$
There is thus a testable prediction of S-duality and moreover Sen was able to 
explicitly present the appropriate self-dual 2-form on the Atiyah-Hitchin manifold
thus confirming the prediction in the simplest case of a 2-monopole bound state.
As we have seen the metrics on the higher charge ($k>2$) monopole moduli spaces
${\cal M}_k^0$ are not known, so there is no hope of a similar explicit
construction of the predicted Sen forms. However, the existence of these
forms can be answered by a study of the appropriate cohomology of ${\cal M}_k^0$
and the predictions of S-duality have essentially been confirmed in this
way by Segal \& Selby\cite{SS}, making use of Donaldson's rational map
description of the monopole moduli space. An alternative confirmation
of the existence of the Sen forms has been given by Porrati\cite{Po}.

As we have seen in the last Section there are simplifying special cases
for higher rank gauge groups where the metric is known explicitly. 
For $G=SU(N)$ with maximal symmetry breaking the metric  on the moduli space 
of charge $(1,1,...,1)$ monopoles is known. The appropriate Sen forms
for the simplest case of $N=3$ were presented by Lee, Weinberg \& Yi\cite{LWY1}
and also Gauntlett \& Lowe\cite{GL} and for the case of general $N$ by
Gibbons\cite{Gi1}.

For classical monopoles all configurations of charge $k$ have equal
energy, so there are no special configurations from this point of view.
However, in the quantum case the Sen form determines a probability density
on the classical monopole moduli space which is peaked over certain sets
of classical monopole solutions. In the work of Segal \& Selby\cite{SS}
certain low-dimensional cycles in ${\cal M}_k^0$ are also important.
We have seen in Section 3 that there are special configurations of monopoles,
such as those with Platonic symmetry or extra zeros of the Higgs
field and it would be interesting if these were the ones of interest in the
above context. By constructing a Morse function on ${\cal M}_k^0$
certain distinguished cycles can be found which include these special
configurations\cite{HMS} but more work in this direction is required.

Turning now to the case of ${\cal N}=2$ supersymmetric theories the
celebrated work of Seiberg \& Witten\cite{SW1} allows the computation
of the nonperturbative low energy effective action by application
of a version of duality, but this time which describes how the theory
varies as a function of the expectation value of the Higgs field.
The important object is an algebraic curve of genus $N-1$ for gauge
group $SU(N)$\cite{SW1,KLYT,AF}. A special differential exists
such that the spectrum of BPS states is given by integration of this
differential over the $2(N-1)$ one-cycles of the curve.
There has been much work, which began with that of Gorskii et al\cite{GKMMM},
to connect these curves to those occurring in certain integrable systems.
However, there is also a connection with monopoles, as follows, but it may be only a mathematical
coincidence. The observation\cite{Su2} is that the Seiberg-Witten curves for gauge
group $SU(N)$ have the same form as the quotient spectral curves of $SU(2)$ 
cyclically symmetric charge
$N$ monopoles. At the time it seemed very strange that the size of the gauge group
was connected to the charge of the monopole but, as we shall discuss briefly below,
exactly the same phenomenon was later found for gauge theories in three dimensions
and a physical explanation discovered in terms of string theory.
It was checked by Chalmers \& Hanany\cite{CH} that extending this correspondence
to the gauge group $SO(2N)$ by considering the quotient spectral curves of 
dihedral monopoles also works and it seems likely that it  extends to the exceptional
groups by considering Platonic monopoles. However, from the point of view of the 
monopole construction it is not so clear how to incorporate the exceptional algebras.
In the case of the affine algebra $A_{k-1}^{(1)}$ there
is a simple ansatz\cite{Su2} to determine the form of the Nahm data for $SU(2)$ cyclic
$k$-monopoles in terms of the generators of the algebra. The problem with the exceptional
algebras is that if one attempts to use the same ansatz then the construction will work
but the charge of the monopoles will be ridiculously high. It seems plausible that the
charges can be reduced to reasonable values by considering principle $su(2)$ subalgebras
but still these charges are higher than the minimal values known to allow Platonic configurations.
For these reasons the connection between the Platonic monopoles discussed in Section 3
and the exceptional algebras is still not understood.

Finally, let us discuss the case of the Coulomb branch of ${\cal N}=4$ supersymmetric 
theories in three dimensions\cite{SW2,CH}.
These are the latest developments in Seiberg-Witten theory but appear to be those in which
a direct and interesting connection with results in monopole theory can be made.

The ${\cal N}=4$ supersymmetric theory in three dimensions can be obtained from a dimensional
reduction of the  ${\cal N}=1$ supersymmetric theory in six dimensions. The theory therefore
contains three Higgs fields and, since the Higgs potential is the trace of the commutators of these
fields, the vacuum configurations of Higgs fields are where they all lie in the Cartan
subalgebra of the gauge group $G.$ Consider the case $G=SU(N)$, 
then the moduli space of the Higgs vacuum is
$3(N-1)$-dimensional. In addition, for maximal symmetry breaking,
 there are $N-1$ photons
which in three dimensions are dual to $N-1$ scalars. So in all, the classical
moduli space of vacua is the $4(N-1)$-dimensional space of these massless scalars.
The fact that there is ${\cal N}=4$ supersymmetry means that the low energy effective
action can be understood in terms of the metric on this manifold, which is
 hyperk\"ahler. Classically the metric is flat but there are both perturbative
and instanton corrections to this. Note that since the theory is in three dimensions
then the classical instantons are actually the BPS monopoles themselves.

Using ideas of duality Seiberg \& Witten\cite{SW2} computed the quantum metric in the
$SU(2)$ case and found it to be the Atiyah-Hitchin metric ie. the classical metric on the moduli
space ${\cal M}_2^0$ of $SU(2)$ centred 2-monopoles. The generalization of the quantum metric 
to gauge group $SU(N)$ was conjectured by Chalmers \& Hanany\cite{CH}
 to be the metric on ${\cal M}_N^0$ ie. the classical metric on the moduli space of centred $SU(2)$
monopoles of charge $N.$ An explanation of this intriguing identification in terms of
string theory has been given by Hanany \& Witten\cite{HW} and consists of considering
certain configurations of fivebranes and threebranes in type IIB superstring theory in ten 
dimensions.
 
The obvious question for monopole theorists is how these conjectured appearances
of multi-monopole moduli spaces can be verified. For the quantum theory in three dimensions 
the perturbative loop corrections can be computed as can the non-perturbative instanton corrections,
at least for low instanton numbers. Indeed for the $SU(2)$ case Dorey et al\cite{DKMTV}
have computed the perturbative and one-instanton corrections and shown that they agree
with the asymptotic form of the Atiyah-Hitchin metric plus the leading order exponential
correction. This is enough to verify the $SU(2)$ result.
However, recall from the discussion in Section 4 that the multi-monopole
moduli space metric is not known on ${\cal M}_k^0$ for $k>2.$ Thus even if the computations
of the quantum metric are performed there is no known result for comparison. 

The ball is now in the court of the monopole theorist to provide some results to which
instanton calculations can be compared. Excluding the Atiyah-Hitchin case (and the
known Atiyah-Hitchin submanifolds\cite{Bi2,HS5}) there is only one case in which the
exact monopole metric is known even on any geodesic submanifold of ${\cal M}_k^0.$
This is the metric on the moduli space of tetrahedrally symmetric charge four monopoles\cite{BS}.
The correspondence between the scalar fields in the quantum gauge theory and the moduli
of the monopoles is that the vacuum expectation values of the three Higgs fields gives
the relative positions of the monopoles and the scalars dual to the photons give the
relative phases. Thus by computation of the perturbative and instanton contributions
in the $SU(4)$ quantum theory where the vacuum expectation values correspond to the vertices
 of a tetrahedron, there is a known result against which the answer can be checked.
If required, the metric on other totally geodesic submanifolds discussed in Section 4,
in which the Nahm data is known in terms of elliptic functions, could also be computed.

Seiberg \& Witten\cite{SW1} also discuss the quantum field theory when massive hypermultiplets
are included. If a single hypermultiplet is included, whose mass vector is
${\bf q}\in \R^3$, they find that the appropriate 4-dimensional hyperk\"ahler metric 
is a one-parameter deformation,
${\cal M}(\vert{\bf q}\vert)$, of the Atiyah-Hitchin metric. This metric was discovered by
Dancer\cite{Da1,Da3} as the hyperk\"ahler quotient of the manifold $M_8^0$ (which is the
centred moduli space of charge two $SU(3)$ monopoles as discussed in the last Section)
by a $U(1)$ subgroup of the $SU(2)$ action. The vector ${\bf q}$ is
the level set of the moment map and the manifold is a deformation of the 
Atiyah-Hitchin manifold in the sense that ${\cal M}(0)$ is the double cover of ${\cal M}_2^0.$

Recently, Houghton\cite{Ho1} has rederived the manifold ${\cal M}(\vert{\bf q}\vert)$
 as another monopole moduli space. It is the moduli space of charge $(2,1)$ monopoles 
in a maximally broken $SU(3)$ theory in which the mass of the $(\ ,1)$ monopole is taken to infinity. 
This infinite mass limit fixes the position of the $(\ ,1)$ monopole which is determined by
the constant vector ${\bf q}.$ An advantage of Houghton's description is
that the asymptotic metric can be computed using a point particle approximation. 
However, it should be possible to go further and compute the exponential corrections
to this asymptotic metric using the Nahm data (\ref{dancer}). This would be a useful
result since a comparison with, at least, the one-instanton correction in the quantum theory should be
possible.

The process of taking massless and infinitely massive limits of monopole
moduli spaces has been investigated within the framework of the hyperk\"ahler quotient
construction by Gibbons \& Rychenkova\cite{GR} and other metrics of relevance to
quantum gauge theories in three dimensions obtained.\\

\nonumsection{Acknowledgements}
\noindent
Over a few years I have benefited from useful interactions with several monopole experts.
In particular I would like to thank Ed Corrigan, Nigel Hitchin, Conor Houghton,
Werner Nahm, Richard Ward and especially Nick Manton. 
I am also grateful to Nick Dorey and Ami Hanany for discussions of their work on duality.

\newpage
\nonumsection{References}
\ 


\vskip -1cm
\noindent
\begin{thebibliography}{000}
\bibitem{Bo} E.B. Bogomolny, Sov. J. Nucl. Phys. 24, 449 (1976).
\bibitem{PS} M.K. Prasad and C.M. Sommerfield, Phys. Rev. Lett. 35, 760 (1975).
\bibitem{Wa1} R.S. Ward, Commun. Math. Phys. 79, 317 (1981).
\bibitem{JZ} B. Julia and A. Zee, Phys. Rev. D 11, 2727 (1975).
\bibitem{AH} M.F. Atiyah and N.J. Hitchin,
\lq{\sl The geometry and dynamics of magnetic monopoles}\rq,
Princeton University Press, 1988.
\bibitem{Ma4} N.S. Manton, Nucl. Phys. B 126, 525 (1977).
\bibitem{We1} E.J. Weinberg, Phys. Rev. D 20, 936 (1979).
\bibitem{CG} E. Corrigan and P. Goddard, Commun. Math. Phys. 80, 575 (1981).
\bibitem{Do} S.K. Donaldson, Commun. Math. Phys. 96, 387 (1984).
\bibitem{JT} A. Jaffe and C. Taubes, \lq{\sl Vortices and
monopoles}\rq, Boston, Birkh\"auser, 1980.
\bibitem{Wa4} R.S. Ward, Phys. Lett. A 61, 81 (1977).
\bibitem{WaW} R.S. Ward and R.O. Wells, \lq{\sl Twistor Geometry and Field Theory}\rq, 
Camnbridge University Press, 1990.
\bibitem{AW} M.F. Atiyah and R.S. Ward, Commun. Math. Phys. 55, 111 (1977).
\bibitem{Ma5} N.S. Manton, Nucl. Phys. B 135, 319 (1978).
\bibitem{CF} E. Corrigan and D.B. Fairlie, Phys. Lett. 67, 69 (1977).
\bibitem{tH} G. t'Hooft, {\sl unpublished}.
\bibitem{Wil} F. Wilczek, in \lq{\sl Quark confinement and field theory}\rq, 
ed. D. Stump and D. Weingarten, John Wiley, New York, 1977.
\bibitem{Hi1} N.J. Hitchin, Commun. Math. Phys. 83, 579 (1982). 
\bibitem{Hi2} N.J. Hitchin, Commun. Math. Phys. 89, 145 (1983).
\bibitem{Pr2} M.K. Prasad, Physica D 1, 167 (1980).
\bibitem{Wa2} R.S. Ward, Phys. Lett. B 102, 136 (1981).
\bibitem{FHP} P. Forg\'acs, Z. Horv\'ath and L. Palla, Phys. Lett. B 99, 232 (1981);
 102, 131 (1981); Ann. of Phys. 136, 371 (1981); Nucl. Phys. B 192, 141 (1981); 229, 77
(1983).
\bibitem{PR} M.K. Prasad and P. Rossi, Phys. Rev. Lett. 46, 806 (1981).
\bibitem{Pr1} M.K. Prasad, Commun. Math. Phys. 80, 137 (1981).
\bibitem{Hu1} J. Hurtubise, Commun. Math. Phys. 92, 195 (1983).
\bibitem{Na} W. Nahm, \lq{\sl The construction of all self-dual
multimonopoles by the ADHM method}\rq, in Monopoles in quantum field
theory, eds. N.S. Craigie, P. Goddard and W. Nahm, World Scientific, 1982.
\bibitem{ADHM} M.F. Atiyah, N.J. Hitchin, V.G. Drinfeld and Yu.I. Manin, Phys. Lett. A 65,
185 (1978).
\bibitem{BPP} S.A. Brown, H. Panagopoulos and  M.K. Prasad, 
Phys. Rev. D 26, 854 (1982); 28, 380 (1983).
\bibitem{Hu2} J. Hurtubise, Commun. Math. Phys. 100, 463 (1985).
\bibitem{HMM} N.J. Hitchin, N.S. Manton and M.K. Murray, Nonlinearity, 8, 661 (1995).
\bibitem{Bi1} R. Bielawski, Ann. Glob. Anal. Geom. 14, 123 (1996).
\bibitem{HS2} C.J. Houghton and P.M. Sutcliffe, Nonlinearity 9, 385 (1996).
\bibitem{Ja} S. Jarvis, \lq{\sl A rational map for Euclidean monopoles via
radial scattering}\rq, Oxford preprint (1996).
\bibitem{BTC} E. Braaten, S. Townsend and L. Carson, Phys. Lett. B 235, 147 (1990).
\bibitem{Kl} F. Klein, \lq{\sl Lectures on the icosahedron}\rq,
London, Kegan Paul, 1913.
\bibitem{HS1} C.J. Houghton and P.M. Sutcliffe, Commun. Math. Phys. 180, 343 (1996).
\bibitem{HS3} C.J. Houghton and P.M. Sutcliffe, Nucl. Phys. B 464, 59 (1996).
\bibitem{At2} M.F. Atiyah, \lq{\sl Magnetic monopoles in hyperbolic spaces}\rq, 
in M.F. Atiyah Collected Works, vol. 5, 579.
\bibitem{JNR} R. Jackiw, C. Nohl and C. Rebbi, Phys. Rev. D 15, 1642 (1977).
\bibitem{Su5} P.M. Sutcliffe,  Phys. Lett. B 376, 103 (1996).
\bibitem{HMS} C.J. Houghton, N.S. Manton and P.M. Sutcliffe, \lq{\sl Rational maps,
monopoles and Skyrmions}\rq, hep-th/9705151.
\bibitem{Ma1} N.S. Manton, Phys. Lett. B 110, 54 (1982).
\bibitem{Sa} T.M. Samols, Commun. Math. Phys. 145, 149 (1992).
\bibitem{Ru} P.J. Ruback, Nucl. Phys. B 296, 669 (1988).
\bibitem{Str} I.A.B. Strachan, J. Math. Phys. 33, 102 (1992).
\bibitem{Wa5} R.S. Ward, Phys. Lett. B 158, 424 (1985).
\bibitem{Le} R.A. Leese, Nucl. Phys. B 344, 33 (1990).
\bibitem{St} D. Stuart, Commun. Math. Phys. 166, 149 (1994).
\bibitem{HKLR} N.J. Hitchin, A. Karlhede, U. Lindstr\"om and M. Ro\v cek,
 Commun. Math. Phys. 108, 535 (1987).
\bibitem{LPZ} R.A. Leese, M. Peyrard and W.J. Zakrzewski, Nonlinearity 3, 387 (1990).
\bibitem{Su6} P.M. Sutcliffe, Nonlinearity 4, 1109 (1991).
\bibitem {BM} L. Bates and R. Montgomery, Commun. Math. Phys. 118, 635 (1988).
\bibitem{HS5} C.J. Houghton and P.M. Sutcliffe, Nonlinearity 9, 1609 (1996).
\bibitem{Bi2} R. Bielawski, Nonlinearity 9, 1463 (1996).
\bibitem{BS} H.W. Braden and P.M. Sutcliffe, Phys. Lett. B 391, 366 (1997).
\bibitem{Nak1} H. Nakajima, \lq{\sl  Monopoles and Nahm's equations
}\rq, in Sanda 1990, Proceedings, Einstein metrics and
 Yang-Mills connections.
\bibitem{Su1} P.M. Sutcliffe,  Phys. Lett. B 357, 335 (1995).
\bibitem{KPZ} A. Kudryavtsev, B. Piette and W.J. Zakrzewski,
Phys. Lett. A 180, 119 (1993).
\bibitem{Su2} P.M. Sutcliffe, Phys. Lett. B 381, 129 (1996).
\bibitem{Wa3} R.S. Ward, J. Phys. A 20, 2679 (1987).
\bibitem{Su3} P.M. Sutcliffe, J. Phys. A 29, 5187 (1996).
\bibitem{Su4} P.M. Sutcliffe, \lq{\sl Cyclic Monopoles}\rq, hep-th/9610030, to appear in Nucl.
Phys. B.
\bibitem{GM2} G.W. Gibbons and N.S. Manton, Phys. Lett. B 356, 32 (1995).
\bibitem{Ma6} N.S. Manton, Phys. Lett. B 154, 397 (1985); (E) B 157, 475 (1985).
\bibitem{GNO} P. Goddard, J. Nuyts and D. Olive, Nucl. Phys. B 125, 1, (1977).
\bibitem{We2} E.J. Weinberg, Nucl. Phys. B 167, 500 (1980).
\bibitem{Wa6} R.S. Ward, Commun. Math. Phys. 86, 437 (1982).
\bibitem{HM} J. Hurtubise and M.K. Murray, Commun. Math. Phys. 122, 35 (1989).
\bibitem{Mu1} M.K. Murray, Commun. Math. Phys. 125, 661 (1989).
\bibitem{Co} S. Connell, \lq{\sl The dynamics of the $SU(3)$ $(1,1)$ magnetic
monopole}\rq. PhD Thesis. The Flinders University of South Australia, 1991.
\bibitem{LWY1} K. Lee, E.J. Weinberg and P. Yi, Phys. Lett. B 376, 97 (1996).
\bibitem{GL} J.P. Gauntlett and D.A. Lowe, Nucl. Phys. B 472, 194 (1996).
\bibitem{LWY2} K. Lee, E.J. Weinberg and P. Yi, Phys. Rev. D 54, 1633 (1996).
\bibitem{Mu2} M.K. Murray, \lq{\sl A note on the $(1,1,...,1)$ monopole metric}\rq,
hep-th/9605054.
\bibitem{NT} M. Takahasi, PhD Thesis, University of Tokyo.
\bibitem{GR} G.W. Gibbons and P. Rychenkova, \lq{\sl Hyperk\"ahler quotient
 construction of BPS monopole moduli spaces}\rq, hep-th/9608085.
\bibitem{IR} I.T. Ivanov and M. Ro\v cek, Commun. Math. Phys. 182, 291 (1996).
\bibitem{Ch1} G. Chalmers, \lq{\sl Multi-monopole moduli spaces for $SU(N)$
gauge groups}\rq, hep-th/9605182.
\bibitem{BW} F.A. Bais and D. Wilkinson, Phys. Rev. D 19, 2410 (1979).
\bibitem{LS} A.N. Leznov and M.V. Saveliev, Lett. Math. Phys. 3, 489 (1979); Commun. Math. Phys.
74, 111 (1980).
\bibitem{GGO} N. Ganoulis, P. Goddard and D. Olive, Nucl. Phys. B 205 [FS5] 601 (1982).
\bibitem{Da1} A.S. Dancer, Commun. Math. Phys. 158, 545-568 (1993).
\bibitem{Da2} A.S. Dancer, Nonlinearity 5, 1355 (1992).
\bibitem{Ir} P. Irwin, \lq{\sl $SU(3)$ monopoles and their fields}\rq, hep-th/9704153.
\bibitem{DL1} A.S. Dancer and R.A. Leese, Proc. R. Soc. 440, 421 (1993).
\bibitem{DL2} A.S. Dancer and R.A. Leese, Phys. Lett. B 390, 252 (1997).
\bibitem{LWY3} K. Lee, E.J. Weinberg and P. Yi, Phys. Rev. D 54, 6351 (1996).
\bibitem{HS6} C.J. Houghton and P.M. Sutcliffe, \lq{\sl $SU(N)$ monopoles and Platonic
symmetry}\rq, preprint UKC/IMS/96-70.
\bibitem{MO} C. Montonen and D. Olive, Phys. Lett. B 72, 117 (1977).
\bibitem{Os} H. Osborn, Phys. Lett. B 83, 321 (1979).
\bibitem{Se} A. Sen, Phys. Lett. B 329, 217 (1994).
\bibitem{SS} G. Segal and A. Selby, Commun. Math. Phys. 177, 775 (1996). 
\bibitem{Po} M. Porrati, Phys. Lett. B 377, 67 (1996).
\bibitem{Gi1}  G.W. Gibbons, Phys. Lett. B 382, 53 (1996).
\bibitem{SW1} N. Seiberg and E. Witten, Nucl. Phys. B 426, 19 (1994).
\bibitem{KLYT} A. Klemm, W. Lerche, S. Yankielowicz and
S. Theisen, Phys. Lett. B344, 169 (1995).
\bibitem{AF} P. Argyres and A. Faraggi, Phys. Rev. Lett. 73, 3931 (1995).
\bibitem{GKMMM} A. Gorskii, I. Krichever, A. Marshakov,
A. Mironov and A. Morozov, Phys. Lett. B355, 466 (1995).
\bibitem{CH} G. Chalmers and A. Hanany, Nucl. Phys. B 489, 223 (1997).
\bibitem{SW2} N. Seiberg and E. Witten, \lq{\sl Gauge dynamics and compactification to three
dimensions}\rq, hep-th/9607163.
\bibitem{HW} A. Hanany and E. Witten, Nucl. Phys. B 492, 152 (1997).
\bibitem{DKMTV}  N. Dorey, V.V. Khoze, M.P. Mattis, D. Tong and S. Vandoren,
\lq{\sl Instantons, three-dimensional gauge theory, and the Atiyah-Hitchin manifold}\rq,
hep-th/9703228.
\bibitem{Da3} A.S. Dancer, Quart. J. Math. 45, 463 (1994).
\bibitem{Ho1} C.J. Houghton, \lq{\sl New hyperk\"ahler manifolds by fixing monopoles}\rq,
hep-th/9702161.
\end{thebibliography}
\end{document}